\documentclass[%
aps,
prd,
twocolumn,
superscriptaddress,
preprintnumbers,
nofootinbib,
amsmath,amssymb,
]{revtex4-2}

\usepackage{style}
\usepackage{aas_macros}

\begin{document}

\preprint{Imperial--TP--2026--AM--01}
\preprint{DESY-26-070}

\title{Detecting Gravitational-Wave Anisotropies with Simulation-Based Inference}

\author{Anna-Malin Lemke\,\orcidlink{0009-0005-3568-3336}}
\affiliation{Deutsches Elektronen-Synchrotron DESY, Notkestr. 85, 22607 Hamburg, Germany}
\email{anna-malin.lemke@desy.de}

\author{Andrea Mitridate\,\orcidlink{0000-0003-2898-5844}}
\affiliation{Abdus Salam Centre for Theoretical Physics, Imperial College, London, SW7 2AZ, UK}
\email{a.mitridate@imperial.ac.uk}

\author{Thomas Konstandin\,\orcidlink{0000-0002-2492-7930}}
\affiliation{Deutsches Elektronen-Synchrotron DESY, Notkestr. 85, 22607 Hamburg, Germany}

\author{Mauro Pieroni\,\orcidlink{0000-0003-0665-266X}}
\affiliation{
Instituto de Estructura de la Materia (IEM), CSIC, Serrano 121, 28006 Madrid, Spain}

\author{James Alvey\,\orcidlink{0000-0003-2020-0803}}
\affiliation{Kavli Institute for Cosmology Cambridge, Madingley Road, Cambridge CB3 0HA, United Kingdom}
\affiliation{Institute of Astronomy, University of Cambridge, Madingley Road, Cambridge CB3 0HA, United Kingdom}

\begin{abstract}
\noindent Over the last five years, multiple Pulsar Timing Array (PTA) collaborations have reported mounting evidence for a gravitational-wave background (GWB) at nanohertz frequencies. Measuring anisotropies in the sky distribution of the GWB power is one of the most promising ways to identify and characterize its source. These anisotropies are expected to manifest as deviations from the Hellings-Downs (HD) correlations between the timing residuals of different pulsars. Current search strategies include Bayesian methods, which model anisotropies in the timing residuals likelihood, and faster frequentist approaches, which construct correlation estimators from timing residuals and use these to test the isotropic assumption. However, frequentist methods rely on the assumption that correlation estimators are Gaussian-distributed, an assumption that is not justified and that -- as we will show -- severely limits detection sensitivity. In this work, we present a Simulation-Based Inference (SBI) framework that replaces the analytic Gaussian likelihood used in frequentist searches with a neural network classifier trained on synthetic data.  This approach captures the non-Gaussian structure of the data and significantly improves performance. Specifically, we find that the probability of $3\sigma$ detection increases by approximately \onegain for single-hotspot scenarios and by \twogain for double-hotspot scenarios compared to standard frequentist methods.
\end{abstract}

\maketitle
\tableofcontents

\section{Introduction}
\noindent Pulsar timing arrays (PTAs) operate by monitoring the radio emission from collections of galactic millisecond pulsars. The exceptional rotational stability of millisecond pulsars allows for the measurement of their rotational, astrometric, and binary parameters (as well as other ephemeris parameters) from the arrival times of these radiation pulses. These parameters are then used to construct a \emph{timing model} that can be used to predict future pulses' times of arrival (TOAs)~\cite{Hobbs:2006cd, 2004hpa..book.....L, Taylor:1993an}.

In 1983, Hellings \& Downs suggested that the correlations between the TOA deviations from timing model predictions could be used to detect GW signals that would otherwise be buried under instrumental and pulsar noise. Indeed, they showed that the presence of a gravitational-wave background (GWB) would induce a characteristic correlation pattern in the timing residuals across pulsar pairs, a pattern that depends uniquely on the angular separation between pulsars on the sky~\cite{Hellings:1983fr}.

Forty years after the original Hellings \& Downs (HD) prediction, several PTA collaborations have reported the first evidence for a GWB by measuring -- at different levels of significance -- stochastic TOA perturbations with an interpulsar correlation pattern compatible with the HD predictions~\cite{NANOGrav:2023gor,EPTA:2023fyk,Reardon:2023gzh,Xu:2023wog,Miles:2024rjc}. While further data is needed to confirm this discovery, the question of the GWB's origin has already sparked tremendous interest. Indeed, while a population of supermassive black hole binaries (\mbox{SMBHBs}) provides a compelling explanation (see, for example, Refs.~\cite{NANOGrav:2023hfp, EPTA:2023xxk}), alternative primordial sources (such as cosmic strings, phase transitions, scalar-induced GWs, or inflationary gravitational waves) cannot be ruled out at the moment (see, for example, Refs.~\cite{NANOGrav:2023hvm, EPTA:2023xxk}). 

One of the most promising avenues to identify the GWB origin is to search for anisotropies in the sky distribution of its power. Indeed, SMBHBs are expected to produce localized anisotropies corresponding to the locations of bright binaries (see, for example, Refs.~\cite{Taylor:2013esa, Gardiner:2023zzr, Lemke:2024cdu}) within the sensitivity reach of future PTAs (see, for example, Refs.~\cite{Pol:2022sjn, Gardiner:2023zzr, Lemke:2024cdu, Depta:2024ykq}). In contrast, most cosmological sources would generate a nearly isotropic GWB with anisotropies well below the sensitivity of current and future PTAs~\cite{Caprini:2018mtu,LISACosmologyWorkingGroup:2022kbp}. Consequently, any detection of significant anisotropies would provide strong evidence for an astrophysical origin of the signal.

GWB anisotropies manifest in PTA observations by modifying the HD correlations induced in the timing residuals by the GWB~\cite{Mingarelli:2013dsa}. Searches for these deviations from the HD correlation pattern have been carried out using both Bayesian and frequentist techniques. In Bayesian approaches (see, for example, Refs.~\cite{Taylor:2013esa,Taylor:2015udp, NANOGrav:2023tcn}), the measured timing residuals are analyzed to reconstruct the GWB sky map. In frequentist approaches, the timing residuals are first compressed into cross-correlation estimators, which are then used to test the isotropic assumption (see, for example, Refs.~\cite{Pol:2022sjn, NANOGrav:2023tcn}). Frequentist searches are significantly faster than their Bayesian counterparts, enabling extensive simulation campaigns that would be computationally prohibitive otherwise. For instance, this speed advantage allows one to derive robust null distributions by analyzing thousands of mock datasets that contain isotropic GWB signals, or to explore the sensitivity of anisotropic searches to different SMBHB population parameters (see, for example, Refs.~\cite{Lemke:2024cdu,Konstandin:2025ifn}).

However, the speed of frequentist anisotropy searches comes at the price of several assumptions that limit their sensitivity. A critical assumption is that cross-correlation estimators extracted from the timing residuals are Gaussian-distributed. Indeed, as already pointed out in Ref.~\cite{Hazboun:2023tiq}, these cross-correlation estimators do not follow a Gaussian distribution. The authors of Ref.~\cite{Hazboun:2023tiq} derived analytical marginalized distributions for individual cross-correlation coefficients, but the full $N_{\rm pair}$-dimensional joint distribution (with $N_{\rm pair} \sim 7000$ for current PTA datasets) remains challenging to derive analytically.

In this work, we address this limitation by developing a simulation-based inference (SBI) framework that circumvents the need for an analytical likelihood. Instead of relying on the Gaussian approximation, we train a graph neural network classifier on synthetic data to distinguish between isotropic and anisotropic GWBs. This data-driven approach naturally captures the non-Gaussian and correlated structure of the cross-correlation estimators, significantly improving anisotropy detection sensitivity. Specifically, the probability of $3\sigma$ detection for a single GWB hotspot increases by approximately \onegain compared to standard frequentist methods, while for two hotspots the detection probability increases by approximately \twogain.

This paper is organized as follows. In Sec.~\ref{sec:review}, we review current strategies for anisotropy searches in PTA observations. In Sec.~\ref{sec:limitations}, we discuss the limitations of these frequentist search strategies. In Sec.~\ref{sec:sbi}, we introduce the SBI framework developed in this work. Our main results, together with a validation of our newly proposed search pipeline, are discussed in Sec.~\ref{sec:results}. Finally, we conclude in Sec.~\ref{sec:conclusions}.           

\section{PTA searches for GWB anisotropies}\label{sec:review}
\noindent In this section, we briefly review how GWB anisotropies manifest in PTA data, as well as the Bayesian (Sec.~\ref{subsec:bayesian}) and frequentist (Sec.~\ref{subsec:frequentist}) strategies used to detect them. This review serves mainly to establish notation and set the stage for our discussion; for a more detailed discussion, see, for example, Refs.~\cite{Pol:2022sjn, Taylor:2013esa,NANOGrav:2023tcn, Konstandin:2025ifn}.

The metric perturbation, $h_{ij}(t,{\bf x})$, associated with a GWB produced by far-away sources can be written as a superposition of plane-waves:
\begin{equation}\label{eq:gwb_metric}
    h_{ij}(t,\vec{x})=\sum_A\int_{-\infty}^{\infty}\!df\int_{S^2}\!d\hat\Omega\; \tilde h_A(f,\hat\Omega) e^{i 2\pi f(t-\hat\Omega\cdot\vec{x})}e_{ij}^A(\hat{\Omega}),
\end{equation}
where $f$ is the GW frequency, $\hat\Omega$ the direction of propagation of the plane waves, $A=+,\times$ labels the two GW polarizations, $e_{ij}^A$ are the GW polarization tensors, and $\tilde h_A(f,\hat\Omega)$ are two complex functions (one for each GW polarization) satisfying $\tilde h_A^*(f,\hat\Omega)=\tilde h_A(-f,\hat\Omega)$. For a GWB arising from the overlapping signal of a large number of sources, the functions $\tilde h_A(f,\hat \Omega)$ can be treated as Gaussian random variables, fully characterized by the two-point function:\footnote{Here we have further assumed the GWB is unpolarized, stationary, and homogeneous, which gives rise to the factors $\delta_{AA'}$, $\delta(f-f')$, and $\delta(\hat{\Omega}, \hat{\Omega}')$, respectively.}
\begin{equation}\label{eq:covariance}
    \langle \tilde h_A^*(f,\hat\Omega)\tilde h_{A'}(f',\hat\Omega')\rangle = \delta_{AA'}\delta(f-f')\delta(\hat{\Omega},\hat{\Omega}')H(f,\hat\Omega)\,,
\end{equation}
where the function $H(f,\hat{\Omega})$ can be factorized as $H(f,\hat{\Omega})=H(f)P(\hat{\Omega},f)$, with $H(f)$ being the GWB power spectrum and $P(\hat{\Omega},f)$ the (normalized) sky map describing the distribution of GWB power on the sky. One of the main goals of anisotropy searches is to test if the signal observed in PTA data is consistent with the isotropic assumption, i.e. $P(\hat\Omega,f)=1$ (here we have normalized the sky map such that $\int d\hat\Omega \, P(\hat\Omega,f)=4\pi$). 

The stochastic signal produced by the GWB in the timing residuals, $\delta t_a(t)$, can be described by the two point function
\begin{equation}\label{eq:res_corrs}
    \langle \delta t_a(t_i)\delta t_b(t_j)\rangle = \int_{-\infty}^\infty df\; \rho_{ab}(f)\Phi(f)e^{2\pi i f (t_j-t_i)}\,,
\end{equation}
where the indices $a,b$ run over pulsars and $i,j$ run over TOAs, and we have defined the timing residuals power spectral density $\Phi(f)\equiv 2H(f)/(3\pi f^2)$ and the cross-correlations between pulsar pairs, $\rho_{ab}$, as
\begin{equation}\label{eq:orf}
    \rho_{ab}(f)=\frac{3}{2}\sum_A\int_{S^2}\frac{d\hat\Omega}{4\pi}\;R_a^A(f,\hat\Omega)R_b^A(f,\hat\Omega) P(\hat\Omega,f)\,.
\end{equation}
The response function, $R_a^A(f,\hat\Omega)$, for the $a^{\text{th}}$ pulsar in the array is given by 
\begin{equation}\label{eq:response}
    R^A_a(f,\hat\Omega)\equiv F^A_a(\hat\Omega) \left[1-e^{-2\pi i f L_a(1+\hat p_a\cdot\hat\Omega)}\right]\,,
\end{equation}
where
\begin{equation}
    F^A_a(\hat\Omega)\equiv\frac{\hat p_a^i\hat p_a^j}{2(1+\hat\Omega\cdot \hat p_a)}e^A_{ij}(\hat\Omega)\,,
\end{equation}
with $\hat p_a$ being the unit vector pointing from Earth to the $a^{\text{th}}$ pulsar, and $L_a$ the distance from Earth to the $a^{\text{th}}$ pulsar in the array. The first term in the square brackets of Eq.~\eqref{eq:response} corresponds to the ``Earth term'', while the exponential in the brackets is usually called the ``pulsar term''.

For an isotropic GWB, the integral in Eq.~\eqref{eq:orf} does not depend on the power spectrum, and the cross-correlations between pulsars become proportional to the well-known Hellings-Downs (HD) function~\cite{Hellings:1983fr}:
\begin{equation}
    \Gamma_{ab} \equiv \frac{1}{2} \delta_{ab} + \frac{1}{2} - \frac{1}{4} x_{ab} + \frac{3}{2} x_{ab} \ln x_{ab}, 
\end{equation}
where $x_{ab} = ( 1- \hat{p}_a \cdot \hat{p}_b) /2$. Therefore, any search for GWB anisotropies consists of testing whether the GWB-induced signal correlations follow the HD correlation pattern. In the remainder of this section, we will review existing Bayesian and frequentist approaches to this problem. 

\subsection{Bayesian}\label{subsec:bayesian}
\noindent Bayesian search strategies use the full information encoded in the timing residuals by building a likelihood that takes the form:\footnote{In this schematic discussion of the PTA likelihood, we are ignoring the contributions from the timing model parameters, which are typically marginalized over in the full analysis.}
\begin{equation}
    p(\bm{\delta t}|\bm{\eta})=\frac{\exp\left(-\frac{1}{2}\bm{\delta t}^T\bm{K}^{-1}\bm{\delta t}\right)}{\sqrt{{\rm det}(2\pi\bm{K})}},
\end{equation}
where $\bm{K}_{ab}(\bm\eta)=\langle\boldsymbol{\delta t}_a \boldsymbol{\delta t}_b\rangle$ is the covariance matrix of timing residuals, and $\boldsymbol{\delta t}_a$ is a vector containing the measured timing residuals for the $a^{\rm th}$ pulsar. This covariance matrix, in addition to noise sources, will also include the contribution from the GWB defined in Eq.~\eqref{eq:res_corrs}. Bayesian search strategies then derive posterior distributions for the sky map parameters, included in $\bm{\eta}$, by sampling the likelihood using Markov Chain Monte Carlo (MCMC) techniques.

A major downside of Bayesian searches is their computational cost. For the NANOGrav 15-year dataset, a frequency-resolved Bayesian search takes on the order of weeks to complete. This high computational cost makes it challenging both to calibrate the searches and to use them in forecast studies, as both tasks would require running multiple instances of the searches on synthetic data to derive null distributions and estimate detection probabilities. While GPU-accelerated data analysis tools -- such as \texttt{Discovery}~\cite{Vallisneri_nanograv_discovery_2025, discovery} -- could help mitigate this issue, frequentist searches remain considerably faster and more straightforward to calibrate.

\subsection{Frequentist}\label{subsec:frequentist}
\noindent Frequentist searches typically proceed in three steps:

\begin{enumerate}
    \item The timing residuals are compressed into an estimator of the cross-correlation coefficients, $\hat\rho_{ab,k}$, as follows:
        \begin{equation}\label{eq:hat_rho}
            \hat\rho_{ab,k}=\boldsymbol{\delta t}_a^T\cdot{\bm w}_{ab,k}\cdot \boldsymbol{\delta t}_b\,,
        \end{equation}
    where $k$ indexes the frequency bin, and the weights, ${\bm w}_{ab,k}$, are defined such that the estimator is unbiased and has minimal variance (see Sec.~\ref{subsec:data} for more details). 
    \item An estimator for the GWB sky map, $\hat{P}(\hat\Omega,f)$, is constructed by maximizing the following Gaussian likelihood:
    \begin{equation}\label{eq:likelihood}
        \hspace{2.5em}p(\hat{\bm{\rho}}_k|\bm{P}_k)=\frac{\exp[-\frac{1}{2}(\hat{\bm{\rho}}_k-\bm{\mathcal{R}}{\bm{P}}_k)^T\mathbf{\Sigma}_k^{-1}(\hat{\bm{\rho}}_k-\bm{\mathcal{R}}{\bm{P}}_k)]}{\sqrt{\det(2\pi\mathbf{\Sigma}_k)}}\,,
    \end{equation}
    where $\bm\Sigma_{k}$ is the covariance matrix of the cross-correlation estimators, $\bm{\mathcal{R}P}_k$ is the discrete form of the integral in Eq.~\eqref{eq:orf}, with ${\bm P}_k$  being a vector containing the GWB power in each equal-area pixel for the sky map at the $k$-th frequency bin, and the (quadratic) response matrix $\bm{\mathcal{R}}$ given by:
    \begin{equation}\label{eq:antenna_response}
        \mathcal{R}_{p,ab}\equiv\frac{3}{2 N_{\rm pix}}\left[F_{a,p}^+F_{b,p}^++ F_{a,p}^\times F_{b,p}^\times\right]\,,
    \end{equation}
    where $p$ runs over a set of $N_{\rm pix}$ equal-area pixels of the GWB sky map, and we have defined $F_{a,p}^A\equiv F^A_a(\hat\Omega_p)$ with $\hat\Omega_p$ a unit
    vector pointing from the $p^{\rm th}$ pixel to the Earth location. The normalization of $\bm{\mathcal{R}}$ is chosen such that for an isotropic sky we recover the HD correlations, i.e. $\bm{\mathcal{R}P}_k=\bm{\Gamma}$ for $P_{k,p}=1$.\footnote{The pulsar term is ignored in the discretized form of the response, as its contribution to Eq.~\eqref{eq:orf} averages out in the case of an isotropic GWB.}
    
    \item The sky map reconstructed in this way is used to define a detection statistic that quantifies deviations from the isotropic null hypothesis (see discussion below). 
\end{enumerate}

Several possible combinations of map parametrizations and detection statistics can be used in frequentist searches~\cite{Konstandin:2025ifn}. In this work, we compare the classifier to the frequentist results obtained using a radiometer basis, which was shown to be one of the best-performing parametrizations~\cite{Konstandin:2025ifn} for detecting localized GWB hotspots. In this parametrization, the GWB power is assumed to be dominated by a single bright pixel. Under this assumption, rather than reconstructing the GWB power distribution across the entire sky, we evaluate one pixel at a time and determine the power required in each pixel to optimally fit the measured cross-correlations. These optimal pixel values can be derived analytically as
    \begin{equation}
        \hat{\bm{P}}_k={\rm diag}(\bm{M}_k)^{-1}\bm{X}_k,        
    \end{equation}
    where $\bm{M}_k=\bm{\mathcal{R}}^{T}\bm{\Sigma}_k^{-1}\bm{\mathcal{R}}$ is the Fisher information matrix and $\bm{X}_k=\bm{\mathcal{R}}^{T}\bm{\Sigma}^{-1}_k\bm{\rho}_k$ is the ``dirty map". In this parametrization, the inverse of the diagonal elements of the Fisher matrix provides an estimate of the uncertainty associated with each reconstructed pixel value, i.e., $\sigma_{k;p}=({M}_{k;pp})^{-1/2}$.
This radiometer map parametrization is then combined with the Max Radiometer SNR detection statistic, defined as ${\rm SNR}={\rm max}[P_{k;p}/\sigma_{k;p}]$, where the maximum is evaluated across all pixels of the reconstructed map.

\section{Limitations of current techniques}\label{sec:limitations}
\noindent In this section, we discuss some of the assumptions that limit the sensitivity of frequentist anisotropy searches. We focus on limitations that our SBI method is designed to address.

\begin{figure}[t]
\centering
\includegraphics{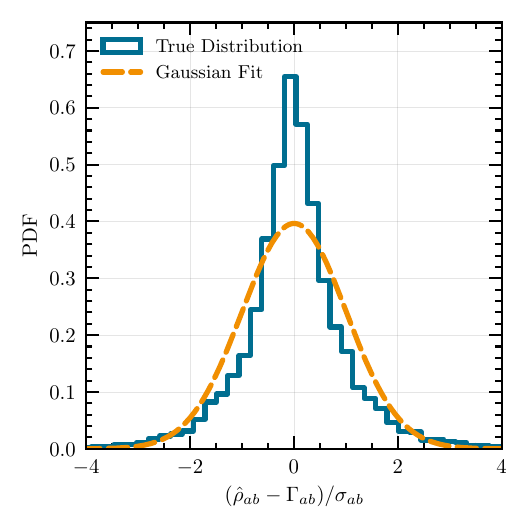}
\caption{Distribution of cross-correlation estimators in the second frequency bin across many isotropic GWB realizations for a single pulsar pair (blue histogram), compared with a Gaussian fit (orange dashed line). The estimators are normalized by subtracting the expected Hellings-Downs value and dividing by their standard deviation, $\sigma_{ab}$, defined as $\sigma_{ab}\equiv \Sigma_{ab,ab}^{1/2}/\Phi_2$, where $\Phi_2$ is the amplitude of the GWB in the second frequency bin. The cross-correlation estimators are normalized by the GWB amplitude such that for an isotropic sky $\langle \hat{\rho}_{ab}\rangle =\Gamma_{ab}$.}
\label{fig:rho_dist}
\end{figure}
\subsection{Non-Gaussianities}
\noindent The likelihood function in Eq.~\eqref{eq:likelihood}, which is used to reconstruct the GWB sky maps, implicitly assumes that the cross-correlation estimators, $\hat{\bm\rho}_k$, follow a Gaussian distribution. However, this assumption does not hold true. This can be seen in Fig.~\ref{fig:rho_dist}, where we show the distribution of one of these cross-correlation estimators across many realizations of an isotropic GWB, and compare it with a Gaussian fit. From this figure, it is clear that the marginalized distribution of $\hat\rho$ deviates from a Gaussian distribution, and follows instead a generalized $\chi^2$ distribution, in agreement with the analytical results derived in Ref.~\cite{Hazboun:2023tiq}. While in this figure we single out a single pulsar pair, similar deviations from Gaussianity are observed for all pulsar pairs in the dataset. 

While Ref.~\cite{Hazboun:2023tiq} provides analytical marginalized distributions for individual cross-correlation coefficients, deriving the full $N_{\rm pair}$-dimensional joint distribution analytically remains challenging. The SBI framework introduced in the next section sidesteps this difficulty by training a classifier to distinguish between isotropic and anisotropic GWBs directly from mock cross-correlation estimators. This approach captures the non-Gaussian, correlated structure of the data without requiring an analytical likelihood.

\subsection{Cross-correlation covariance}
\noindent The covariance matrix, $\boldsymbol{\Sigma}_k$, entering the likelihood of Eq.~\eqref{eq:likelihood} depends on the pulsar cross-correlations. However, these quantities depend on the GWB sky map, which is what we are trying to estimate in an anisotropy search. In current searches, HD cross-correlations are assumed when constructing the covariance matrix, but this is an approximation whose impact on the final results is difficult to quantify. 

The SBI approach sidesteps this problem by not requiring an explicitly specified covariance matrix for the cross-correlations. The covariance structure is naturally encoded in the training data through our forward modeling process: each simulated dataset—isotropic or anisotropic—produces cross-correlation estimators whose covariances reflect the underlying sky map. The classifier learns to distinguish between these cases by capturing the statistical structure of the estimators—including their means, covariances, and higher-order non-Gaussian features.

\subsection{Intermediate map reconstruction}
\noindent In classical frequentist searches, once a sky map is obtained by maximizing the likelihood in Eq.~\eqref{eq:likelihood}, a detection statistic must be constructed to quantify the deviation from isotropy contained in the data. However, if we are only interested in detecting deviations from isotropy, this intermediate map-making step is unnecessary, as the optimal test statistic can be constructed directly from cross-correlations. Moreover, this approach introduces several arbitrary choices, such as the parametrization of the GWB sky map and the form of the detection statistic, with no guarantee that any particular choice represents the optimal test for anisotropy.

As we will discuss in the next section, with the SBI method proposed in this work, we can bypass these intermediate steps and construct a classifier that operates directly on the cross-correlation coefficients and learns a detection statistic from data.

\section{The SBI approach}\label{sec:sbi}

\noindent As just discussed in Sec.~\ref{sec:limitations}, classical frequentist searches rely on intermediate map reconstruction and analytic Gaussian approximations to formulate a test statistic for anisotropy. However, if the primary goal is to detect a deviation from isotropy (i.e., hypothesis testing), this intermediate map-making step is unnecessary. Furthermore, by the Neyman-Pearson lemma, the optimal detection statistic for distinguishing between two competing hypotheses -- an isotropic GWB (hypothesis $\mathcal{H}_0$) and an anisotropic GWB (hypothesis $\mathcal{H}_1$) -- is the likelihood ratio or, in a Bayesian framework, the Bayes factor~\cite{Neyman:1933wgr}.

Traditionally, computing the Bayes factor requires evaluating the marginal likelihood of the data under both models. For PTA cross-correlation estimators, evaluating this integral is intractable because the true joint probability distribution is highly complex, correlated, and non-Gaussian (as shown in Fig.~\ref{fig:rho_dist}). Standard Bayesian searches bypass this by performing computationally expensive MCMC sampling directly on the timing residuals, while frequentist searches compromise by assuming an analytic, but inaccurate, Gaussian likelihood for the cross-correlation estimators. 

To overcome these limitations, we apply ideas from Simulation-Based Inference. SBI is now an extensively utilized framework across astrophysics and cosmology (and beyond); see Ref.~\cite{Cranmer:2019eaq} for a review and Refs.~\cite{Shih:2023jme, Vallisneri:2024xfk, Lai:2025xov, Laal:2024trp} for applications to PTA data analysis. While most of these applications have been in the context of Bayesian parameter estimation, Ref.~\cite{Jeffrey:2023stk} recently demonstrated that Bayesian model comparison can be entirely recast as a classification optimization problem (see also~\cite{AnauMontel:2024flo}). In particular, if a neural network is trained to classify simulated data as belonging to either $\mathcal{H}_0$ or $\mathcal{H}_1$, the network asymptotically learns the density ratio between the two data-generating distributions. By utilizing specific loss functions during training, the output of the classifier directly provides an amortized estimate of the Bayes factor (see App.~\ref{app:evidence_networks} for more details). 

In this work, we apply this methodology to the PTA anisotropy search. Rather than specifying an explicit, approximate analytic likelihood, we implicitly define the true distributions via forward modeling. We then train a neural network classifier directly on the cross-correlation estimators to distinguish between mock isotropic and anisotropic GWB signals. This approach yields several major advantages:
\begin{enumerate}
    \item \textbf{Optimality:} It directly estimates the Bayes factor, learning the optimal non-linear detection statistic directly from the data.
    \item \textbf{Accuracy:} It captures the non-Gaussian structure and covariance of the cross-correlation estimators without requiring analytical derivations.
    \item \textbf{Efficiency:} It bypasses both the arbitrary choices inherent in classical map reconstruction and the massive computational cost of MCMC, allowing for near-instantaneous inference once the network is trained.
\end{enumerate}
The exact test statistic, or Bayes factor, that is learnt depends on the forward modeling assumptions in the various data generation pipelines (e.g. regarding the amplitude, number, and distribution of anisotropies). Below, we discuss our specific choices for these assumptions and, therefore, the hypothesis test we are implicitly carrying out. 

\subsection{Forward modeling}\label{subsec:data}
\noindent To generate the mock cross-correlations used to train and test our classifier, we follow as closely as possible the procedure that would be used to derive cross-correlation estimators from real PTA timing residuals. Specifically, following Ref.~\cite{Konstandin:2025ifn}, we proceed in three steps:
\begin{enumerate}
\item \textbf{Timing residuals generation} (Sec.~\ref{subsubsec:residuals}): We generate mock timing residuals in which we inject white and intrinsic red noise, as well as GWB signals.
\item \textbf{Noise parameter inference} (validation dataset only, Sec.~\ref{subsubsec:noise_run}): For a subset of the mock data, used later for validation, we perform a Bayesian run (which models the GWB as common uncorrelated red noise) to obtain posterior distributions for noise and GWB parameters.
\item \textbf{Cross-correlation estimation} (Sec.~\ref{subsubsec:cross_corr}): We construct the cross-correlation estimators from the timing residuals.
\end{enumerate}
In the remainder of this section, we provide more details on each of these three steps.

\subsubsection{Timing residuals generation}\label{subsubsec:residuals}
\noindent We start by generating mock timing residuals for each of the 120 pulsars observed by regional PTAs comprising the International Pulsar Timing Arrays (IPTA). In doing so, we adopt the sky positions, observation times, and measured noise properties from the most recent data releases from each of these collaborations~\cite{EPTA:2023akd,EPTA:2023sfo,  Miles:2024rjc, NANOGrav:2023hde, Zic:2023gta}.
We assume that the residuals receive contributions from three processes: white noise, intrinsic red noise (IRN), and GWB (which we model as an isotropic component plus anisotropic contributions). Therefore, we model the timing residuals as
\begin{equation}
    \boldsymbol{\delta t}= \boldsymbol{\delta t}_{\rm WN}+\boldsymbol{\delta t}_{\rm IRN}+\boldsymbol{\delta t}_{\rm GWB}+\boldsymbol{\delta t}_{\rm CW}\,,
\end{equation}
where $\boldsymbol{\delta t}_{\rm WN}$ represents the white noise contribution, $\boldsymbol{\delta t}_{\rm IRN}$ the intrinsic red noise, $\boldsymbol{\delta t}_{\rm GWB}$ the contribution from the isotropic component of the GWB, and $\boldsymbol{\delta t}_{\rm CW}$ the contribution from the anisotropic component that we model as a sum of continuous wave (CW) sources. 

\emph{\textbf{White noise}} --
To reduce the dataset to a computationally manageable size, we construct epoch-averaged TOAs by combining all observations of a pulsar at different radio frequencies within a given PTA and observational epoch into a single effective observation.
We compute the effective white noise of the averaged observation, $\sigma_{\rm TOA}$, from the full white noise matrix, $\bm{N}$, which contains measurement uncertainties, additional white noise contributions (EFAC and EQUAD), and pulse jitter for all individual TOAs within a given epoch~\cite{justin_ellis_2017_251456}:\footnote{Notice that this epoch-averaging procedure tends to underestimate the white-noise contribution to the timing residuals. While inflating TOA uncertainties could mitigate this effect, we choose not to do so because the classical method used as a reference in this work employs the same approach. In this way, while the overall sensitivity scale reported here may be overestimated, the relative performance comparison with the classical method remains unbiased.}
\begin{equation}\label{eq:eff_wn}
    \sigma_{\rm TOA}^2 = \Big[\sum_{i,j}\left(N^{-1}\right)_{ij}\Big]^{-1}\,.
\end{equation}
The white noise contribution to the timing residuals is then modeled as a Gaussian process, whose two-point function is given by:
\begin{equation}
    \langle\delta t_{{\rm WN},i}\delta t_{{\rm WN},j}\rangle = \delta_{ij}\sigma_{\rm TOA}^2\,,
\end{equation}
where $i$ and $j$ index the TOAs of a given pulsar. For each observation, we then simulate white noise by drawing from a zero-mean Gaussian distribution with variance given by $\sigma_{\rm TOA}^2$. 

\emph{\textbf{Intrinsic red noise}} --
For all pulsars that show evidence for significant intrinsic red noise according to the noise analyses from the individual PTAs~\cite{EPTA:2023akd, Miles:2024rjc, NANOGrav:2023ctt, Zic:2023gta}, we include this process in our datasets. Following standard conventions, we model IRN as a Gaussian process, for which we can write a single random realization as:
\begin{equation}
    \delta t_{{\rm IRN},i}=\boldsymbol{Fa}=\sum_{j=1}^{N_f}\Big[X_j\sin(2\pi f_jt_i)+Y_j\cos(2\pi f_jt_i)\Big]\,,
\end{equation}
where alternating $X$, $Y$ coefficients make up the Fourier coefficient vector $\boldsymbol{a}$, the design matrix $\boldsymbol{F}$ contains alternating columns of sine and cosine components evaluated at the observation times, $f_j = j/T_{\rm psr}$ where $T_{\rm psr}$ is the individual observation baseline for each pulsar, and we truncate the sum at $N_f=30$ following the convention of Ref.~\cite{NANOGrav:2023gor}. We then generate the IRN contribution to the timing residuals by sampling, for each pulsar, a set of Fourier coefficients from a zero-mean Gaussian distribution with covariance given by:
\begin{equation}
    \langle Y^a_iY^b_j\rangle=\langle X^a_iX^b_j\rangle =\delta_{ij}\delta_{ab}\,\varphi_a(f_i),
\end{equation}
where the IRN power spectral density (PSD) is parametrized as
\begin{equation}
\varphi_a(f)=\frac{A_a^2}{12\pi^2}\left(\frac{f}{{\rm yr}^{-1}}\right)^{-\gamma_a}\frac{{\rm yr}^3}{T_{\rm obs}}\,,
\end{equation}
where $T_{\rm obs}$ is the total observing time, and the amplitude $A_a$ and slope $\gamma_a$ are pulsar-dependent parameters set to the maximum posterior values obtained in Refs.~\cite{EPTA:2023akd, Miles:2024rjc, NANOGrav:2023ctt, Zic:2023gta}.

\emph{\textbf{Isotropic GWB component}} -- 
We model the contribution of the isotropic component of the GWB similarly to IRN, i.e., by decomposing it into Fourier components and sampling the Fourier coefficients from a Gaussian distribution. However, compared to IRN, the Fourier coefficients of different pulsars are now correlated according to the HD curve, such that the two-point function now reads:
\begin{equation}
    \langle Y^a_iY^b_j\rangle=\langle X^a_iX^b_j\rangle =\delta_{ij}\,\Gamma_{ab}\,\Phi_a(f_i)\,,
\end{equation}
where the GWB power spectrum is parametrized as
\begin{equation}
    \Phi(f)=\frac{A_{\rm gwb}^2}{12\pi^2}\left(\frac{f}{{\rm yr}^{-1}}\right)^{-\gamma_{\rm gwb}}{\rm yr}^3\Delta f\,.
\end{equation}
The Fourier components are drawn for a discrete frequency array ranging from $f_{\rm min}=1/(10T_{\rm obs})$ up to $f_{\rm max}=300/T_{\rm obs}$ with spacing $\Delta f=1/(10T_{\rm obs})$. 
The timing residuals are obtained by applying a fast Fourier transform to the Fourier components and then interpolating into the observation times.

\begin{figure}[t]
\centering
\includegraphics{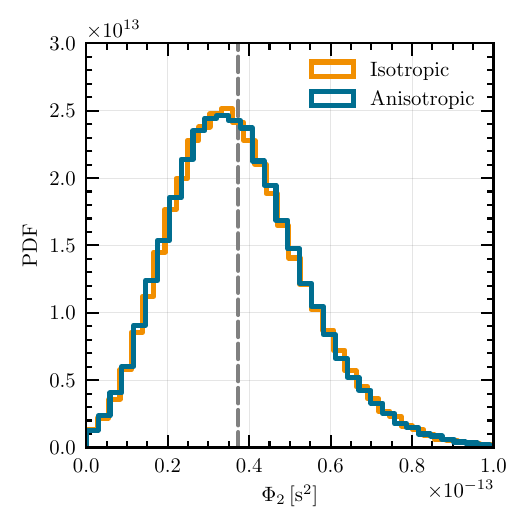}
\caption{Distribution of the recovered timing residuals PSD induced by the GWB in the second frequency bin, $\hat\Phi_2\equiv\hat{\Phi}(f_2)\Delta f$, of the training data. The orange histogram shows the distribution for isotropic GWB realizations, while the blue histogram shows the one for anisotropic realizations containing a varying number of hotspots with varying strength. The vertical dashed line indicates the injected amplitude value, $\Phi_2 = 3.7 \times 10^{-14}\,[\rm{s}^2]$.}
\label{fig:amp_dist}
\end{figure}

\emph{\textbf{Anisotropic GWB component}} --
In this work, we model GWB anisotropies as a superposition of GWB hotspots. This is justified by the expectation that \mbox{SMBHBs} produce localized anisotropies corresponding to the brightest individual binaries in the population\mbox{~\cite{Taylor:2013esa,Gardiner:2023zzr,Lemke:2024cdu}}.
We model these hotspots as CW signals produced by individual SMBHB binaries. The contribution of a CW signal to the timing residuals of the $a^{\text{th}}$ pulsar is given by~\cite{Becsy:2022zbu}:
\begin{widetext}
\begin{equation}
\begin{split}
    \delta t_a(t) = \frac{h_cf_{\scriptscriptstyle\rm GW}^{-3/2}}{2\pi T_{\rm obs}^{1/2}}\Big\{F_a^{+}(\hat{{\Omega}})\Big[&\cos(2\psi)\Big(\sin(\phi_{0}+2\omega t_p)-\sin(\phi_{0}+2\omega t) \Big)\!+\sin(2\psi)\Big(\cos(\phi_{0}+2\omega t_p)
    -\cos(\phi_{0}+2\omega t)\Big)\Big] \\
    +F_a^{\times}(\hat{{\Omega}})\Big[&\sin(2\psi)\Big(\sin(\phi_{0}+2\omega t)-\sin(\phi_{0}+2\omega t_p)\Big)\!+\cos(2\psi)\Big(\cos(\phi_{0}+2\omega t_p)-\cos(\phi_{0}+2\omega t)\Big)\Big]\Big\},
\end{split}
\end{equation}
\end{widetext}
where $h_c$ denotes the characteristic strain of the source associated with the hotspot, $\omega=2\pi f_{\rm GW}$ with $f_{\rm GW}$ the frequency of the GW, $\psi$ is the polarization angle, and $\phi_{0}$ is the initial phase. For each CW source that we inject in the data, $\hat\Omega$ is drawn from a uniform distribution across the sky, while both $\psi$ and $\phi_0$ are drawn from a uniform distribution, $\mathcal{U}(0,2\pi)$. The effect of the pulsar term is encoded in the pulsar time, $t_{p}(t) = (t-L_a(1-\cos \mu))$,  with $\cos\mu=-\hat{{\Omega}}\cdot\hat{{p}}_a$. To reflect our lack of knowledge about the true pulsar distances, we generate new values of $L_a$ for every realization by drawing from a uniform distribution $\mathcal{U}(0.5~\rm{kpc},\,1.5~\rm{kpc})$. 
The characteristic strain for a hotspot contributing to a fraction $x_{\rm hot}$ of the power at frequency $f_{\rm hot}$ is given by
\begin{equation}
    h_c = \sqrt{x_{\rm hot}}\,A_{\rm gwb}\left(\frac{f_{\rm hot}}{\rm {yr}^{-1}}\right)^{-0.5\,(\gamma_{\rm gwb}-3)},
\end{equation}
where, in this work, we set $\gamma_{\rm gwb}=13/3$.
To ensure that our classifier learns to detect anisotropies rather than amplitude variations, we reduce the isotropic GWB spectral density, $\Phi_{\rm iso}(f_{\rm{hot}})$, by the power contributed by the hotspots, ensuring that the total GWB amplitude remains constant across isotropic and anisotropic realizations:
\begin{equation}
    \Phi(f_{\rm{hot}}) 
\to\sqrt{1-x_{\rm hot}}\, \Phi(f_{\rm{hot}}).
\end{equation}
Figure~\ref{fig:amp_dist} demonstrates this explicitly, showing how the distributions of recovered GWB amplitudes overlap for isotropic and anisotropic realizations.\\

After simulating all signal components, the approach closest to a true dataset analysis would include fitting a timing model to each of the pulsars. This, however, significantly increases runtimes and is therefore not feasible for large datasets. Therefore, for the \texttt{training} and \texttt{validation} data sets, we only marginalize over linear deviations from the true timing model parameters when deriving the cross-correlation estimators (see Sec.~\ref{subsubsec:cross_corr} for more details). We have verified that this procedure does not significantly affect the distribution of these estimators~\cite{Konstandin:2025ifn}. Moreover, we also generate a smaller \texttt{test\_real} dataset for which we do perform full timing model fits (using the software package \texttt{pta-replicator}~\cite{pta_replicator}). In Sec.~\ref{sec:results}, we use this realistic validation dataset to explicitly assess the impact of the timing model approximation on classifier performance.

\subsubsection{Noise run}\label{subsubsec:noise_run}
\noindent The weights $\boldsymbol{w}_{ab,k}$ entering the definition of the cross-correlation estimators in Eq.~\eqref{eq:hat_rho} depend on the noise and GWB parameters. In real PTA analyses, posterior distributions for these quantities would be obtained through single-pulsar runs and a preliminary Bayesian analysis modeling the GWB as common uncorrelated red noise (CURN). To marginalize over red noise parameters, the cross-correlations and their uncertainties are then calculated over multiple random draws from the posterior distributions, resulting in what is referred to as the ``noise-marginalized optimal statistic" (NMOS)~\cite{Vigeland:2018ipb}.
However, this procedure is computationally expensive\footnote{As a single analysis run takes multiple hours using the CPU-based \texttt{enterprise} package, or over 10 minutes with the GPU-based \texttt{Discovery}~\cite{Vallisneri_nanograv_discovery_2025, discovery}, it is not feasible to repeat this for the full training dataset containing $\mathcal{O}(10^{6})$ independent simulations.}; therefore, for the \texttt{training} and \texttt{validation} datasets, we set white noise, IRN, and GWB parameters to their injected values. Previous work has shown that fixing white noise, IRN and GWB parameters to their true values does not significantly change the distribution of cross-correlation estimators when considering a large number of simulated datasets~\cite{Vigeland:2018ipb,Konstandin:2025ifn}. Nevertheless, for the smaller \texttt{test\_real} we do perform NMOS runs, so that we can explicitly assess how this approximation affects the performance of our classifier (see Sec.~\ref{sec:results}). 

Specifically, for each timing residuals realization in \texttt{test\_real}, we perform a CURN run to derive the posterior distributions of the intrinsic red noise parameters $(\log_{10}A_a,\gamma_{a})$ and GWB parameters $(\log_{10}A_{\rm gwb},\gamma_{\rm gwb})$. We perform these runs using the GPU-accelerated code \texttt{Discovery}~\cite{Vallisneri_nanograv_discovery_2025, discovery}, assuming a power-law spectral template with 30 and 14 frequency bins for the IRN and GWB, respectively. Then, for each of the timing residuals realizations, we draw 1000 samples of IRN and GWB parameters from these posterior distributions and derive the corresponding cross-correlation estimators according to the procedure outlined in Sec.~\ref{subsubsec:cross_corr}. Therefore, for \texttt{test\_real}, we obtain 1000 sets of these estimators per realization, capturing the uncertainty introduced by imperfect knowledge of the noise and GWB parameters.

\begin{figure*}[th]
    \centering
    \includegraphics[width=\textwidth]{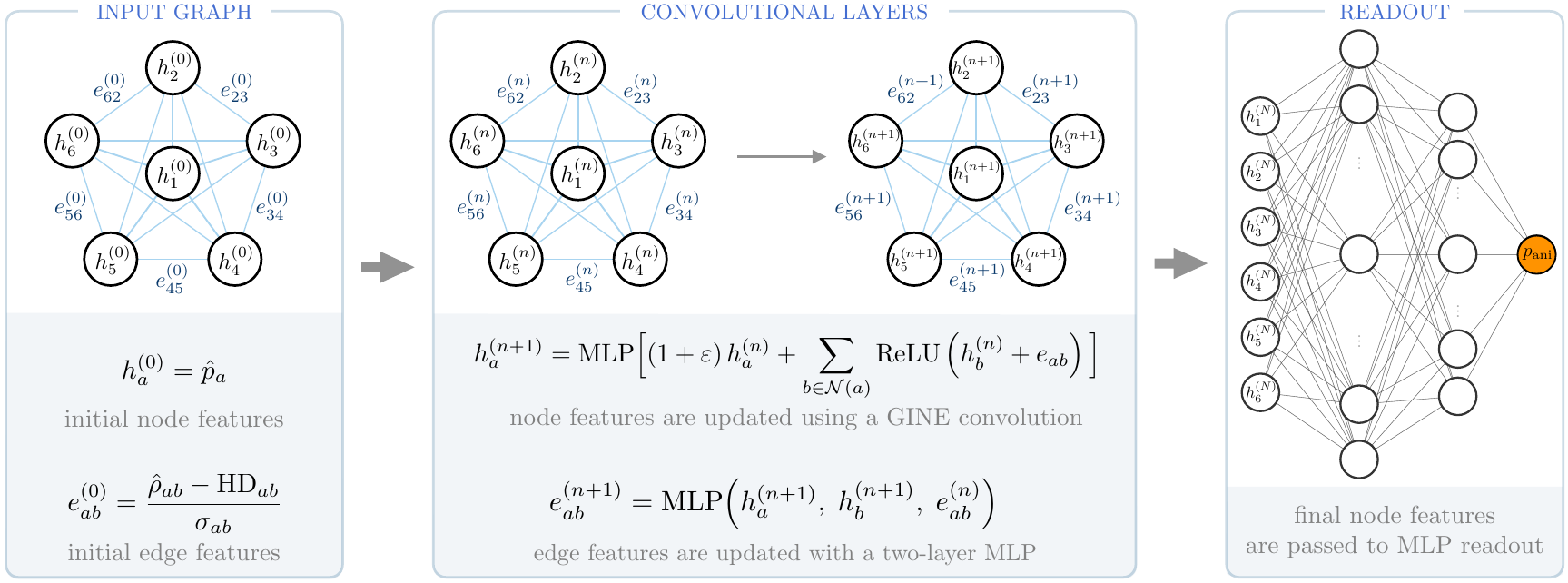}
    \caption{Schematic representation of the GNN classifier used in this work. For clarity, this diagram omits the initial node and edge feature embeddings applied before the convolutional layers and the batch normalization applied after each convolutional layer. See the main text for a complete discussion of the network architecture.}
\label{fig:gnn_architecture}
    \label{fig:roc_comp}
\end{figure*}
\begin{table*}[th]{
\renewcommand{\arraystretch}{2}
    \ra{1.3}
    \begin{center}
    \tabcolsep=0.25cm
    \begin{tabular}{l l l l l l l}
    \hlinewd{1pt}
    \textbf{Dataset} & $\boldsymbol{N_{\rm samples}}$ & \textbf{Noise} & \textbf{TM} & $\boldsymbol{A_{\rm gwb}}$ & $\boldsymbol{n_{\rm hot}}$ & $\boldsymbol{x_{\rm hot}}$ \\
    \hlinewd{1pt}
    $\texttt{training}$        &  $4.6\times10^6$& fixed & marg.  & $[0.25,\,0.5,\,1,\,2,\,4]A_{\rm gwb}^{\rm NG15}$ & $[1,\,2,\ldots,8]$ & $[0.2,\,0.3, \ldots,0.9]$ \\
    $\texttt{validation}$          & $5.7\times10^5$ & fixed & marg. & $[0.25,\,0.5,\,1,\,2,\,4]A_{\rm gwb}^{\rm NG15}$ & $[1,\,2,\ldots,8]$ &  $[0.2,\,0.3, \ldots,0.9]$\\
    $\texttt{test}$          & $5.7\times10^5$ & fixed & marg. & $[0.25,\,0.5,\,1,\,2,\,4]A_{\rm gwb}^{\rm NG15}$ & $[1,\,2,\ldots,8]$ &  $[0.2,\,0.3, \ldots,0.9]$\\
    \texttt{test\_real} &  $2.0\times10^3$& NMOS & fit + marg. & $A_{\rm gwb}^{\rm NG15}$ & 1 & $[0.4,\,0.8]$ \\
    \hlinewd{1pt}
    \end{tabular}
    \end{center}}
\caption{Summary of the properties of the different datasets used in this work. For each dataset, we report: the number of samples ($N_{\rm samples}$); whether IRN and GWB parameters are fixed to their injected values or marginalized via NMOS; whether the timing model (TM) is marginalized over in the OS or also fitted at the timing residuals level; the GWB amplitudes considered (in units of the NANOGrav 15-year amplitude, $A_{\rm gwb}^{\rm NG15}=10^{-14.67}$); the number of hotspots ($n_{\rm hot}$); and the fractional power in hotspots ($x_{\rm hot}$). For all the datasets, samples are split 50-50 between isotropic and anisotropic realizations, with equal representation across all combinations of $A_{\rm gwb}$, $n_{\rm hot}$, and $x_{\rm hot}$ values.}
\label{tab:datasets}
\end{table*}
\subsubsection{Cross-correlation estimation}\label{subsubsec:cross_corr}
\noindent Finally, for each set of mock timing residuals, we use the software package \texttt{DEFIANT} to construct the cross-correlation estimators given in Eq.~\eqref{eq:hat_rho}, where the weights, $\boldsymbol{w}_{ab,k}$, are given by~\cite{Gersbach:2024hcc, Gersbach:2025mhj}: 
\begin{equation}
    \bm{w}_{ab,k}=\frac{\bm{P}_a^{-1}\cdot\skew{5}\tilde{\bm{S}}_{ab,k}\cdot\bm{P}_b^{-1}}{{\rm tr}\left[ \bm{P}_a^{-1}\cdot\skew{5}\tilde{\bm{S}}_{ab,k}\cdot\bm{P}_b^{-1}\cdot\skew{5}\tilde{\bm S}'_{ab,k}\right]}\,.
\end{equation}
The cross-covariance matrix is defined as $\skew{5}\tilde{\bm{S}}'_{ab,k}\equiv \bm{F}_a{\bm{\Phi}}\bm{F}_b^T/\Phi(f_k)$ and $\skew{5}\tilde{\bm{S}}_{ab,k}\equiv \bm{F}_a\tilde{\bm{\phi}}_k\bm{F}_b^T$, where $\tilde{\bm \phi}_k$ is a frequency selector of the form 
\begin{equation}
    \begin{split}
    \tilde{\bm\phi}_1&={\rm diag}(1,1,0,0,\ldots,0,0),\\
    &\vdots\\
    \tilde{\bm\phi}_{N_f}&={\rm diag}(0,0,0,0,\ldots,1,1)\,.
    \end{split}
\end{equation}
The elements of the auto-covariance matrix of pulsar $a$ are given by:
\begin{equation}
    P_{a,ij}\equiv D_{a,ij}+F_{a,ik}(\Phi+\varphi_a)_{kk'}F_{a,jk'}\,,
\end{equation}
where the indices $i$ and $j$ run over the TOAs and $k$ runs over the frequency bins, and we have defined $\bm{\Phi}={\rm diag}(\Phi(f_1), \Phi(f_1), \Phi(f_2), \Phi(f_2), \ldots)\Delta f$, and similarly for $\boldsymbol{\varphi}_a$. 
The matrix $\boldsymbol{D}_a$ is given by
\begin{equation}
    \bm{D}_a = \bm{N}+\bm{M}_a\bm{E}\bm{M}_a^T\,,
\end{equation}
where $\boldsymbol{N}$ is the white noise matrix, and $\boldsymbol{M}$ is an $N_{\rm TOA}\times m$ matrix whose elements are the partial derivatives of the TOAs with respect to the $m$ timing model parameters, evaluated at their best-fit values. The matrix $\boldsymbol{E}$ is diagonal with very large entries (typically $10^{40}$), effectively imposing a flat prior on the timing model parameters. When $\boldsymbol{D}_a$ is inverted, this choice marginalizes over uncertainties in the timing model parameters.

Once a set of \emph{unnormalized} cross-correlation estimators is derived using Eq.~\eqref{eq:hat_rho} in conjunction with the weights given above, we can construct an estimator of the GWB PSD in each frequency bin as~\cite{Gersbach:2025mhj, Gersbach:2024hcc}:
\begin{equation}\label{eq:A_pf_est}
    \hat{\Phi}_k=\frac{\bm{\Gamma}^T\bm{\Sigma}_k^{-1}\hat{\bm{\rho}}_k}{\bm{\Gamma}^T\bm{\Sigma}_k^{-1}\bm{\Gamma}}\,,
\end{equation}
where $\boldsymbol{\Sigma}_{k,ab,cd}\equiv\langle \hat{\rho}_{ab,k} \hat{\rho}_{cd,k}\rangle-\langle\hat{\rho}_{k,ab}\rangle\langle\hat{\rho}_{k,cd}\rangle$ is the covariance matrix of the estimators as derived in~\cite{Gersbach:2024hcc}, which includes contributions from both pulsar noise and GWB self-noise (i.e., cosmic variance in the cross-correlation coefficients~\cite{Konstandin:2024fyo, Domcke:2025esw}). These PSD estimates are then used to normalize the cross-correlation estimators such that their expectation value is given by $\langle \hat\rho_{ab,k}\rangle=\rho_{ab}(f_k)$.

\subsection{Network architecture}
\noindent As detailed in this section, the main task we are trying to solve is a binary classification problem: given a set of cross-correlation estimators, we want to determine whether they are more likely to have been generated by an isotropic GWB ($\mathcal{H}_0$) or an anisotropic GWB containing one or more hotspots ($\mathcal{H}_1$). To solve this problem, we start by representing the data as a graph, which we then feed into a Graph Neural Network (GNN). We have chosen this architecture because of the natural graph structure of the data, and because GNNs are designed to capture complex interactions between nodes and edges, making them well-suited to learn non-trivial patterns in the cross-correlation data that may indicate anisotropy. 

The primary objective of this work is to demonstrate the potential of SBI approaches to anisotropy searches rather than optimizing the network architecture. With this in mind, we implement a minimal GNN with the following architecture:
\begin{enumerate}
    \item \underline{Input graph}: Each observation is represented as a fully-connected graph with one node for each pulsar in the array. To each node of the initial graph, we associate a node attribute, $h_a^{(0)}$, given by the pulsar position in the sky, $h_a^{(0)}=\hat{p}_a$. Similarly, for each edge, we associate an edge attribute, $e_{ab}^{(0)}$, given by the whitened cross-correlation estimator for the pulsar pair connected by that edge. These whitened cross-correlations are obtained by subtracting the Hellings-Downs prediction from the cross-correlation estimator and dividing by the noise variance:
    \begin{equation}
        e_{ab}^{(0)} = \frac{\hat{\rho}_{ab} - \text{HD}_{ab}}{\sigma_{ab}}\,,
    \end{equation}
    where $\sigma_{ab}\equiv \Sigma_{ab,ab}^{1/2}/\hat\Phi$ is the standard deviation of the normalized cross-correlation estimator.
    For our dataset containing $N_p = 120$ pulsars, the graph has 120 nodes and $N_e = 2 \times N_{\rm pair} = N_p (N_p -1 ) = 14,280$ directed edges, where the factor of 2 accounts for representing each undirected edge as two directed edges with opposite orientations. Both the edge and node attributes are standardized to zero mean and unit variance across the entire training data set.
    \item \underline{Edge and node encoders}: A linear projection maps each one-dimensional node feature of the input graph into a higher-dimensional latent space with dimension $d_h=16$. Edge features are also projected into an $d_h$-dimensional latent space using a two-layer Multilayer Perceptron (MLP): the first layer maps the one-dimensional edge feature to dimension $d_h$, followed by a Gaussian Error Linear Units (GELU)~\cite{2016arXiv160608415H} activation; the second layer applies a linear transformation within the $d_h$-dimensional space. This produces $N_e$ edge embeddings of dimension $d_h$.
    \item \underline{Convolutional layers}: Two GINE convolutional layers~\cite{2019arXiv190512265H} update node features by aggregating neighboring node and edge features and using the following update rule:
     \begin{equation}
        \begin{aligned}
            \boldsymbol{h}_a^{(n+1)} = \, & \mathrm{MLP} \left[ (1+\varepsilon)\,\boldsymbol{h}_a^{(n)}\right.\\&\left.+\sum_{b\in\mathcal{N}(a)} \mathrm{ReLU}\!\left(\boldsymbol{h}_b^{(n)} + {\boldsymbol{e}}_{ab} \right)\right].
        \end{aligned}
    \end{equation}
    where $\epsilon$ is a (learnable) scalar parameter that controls the relative weighting of the node’s own embedding versus the aggregated messages from its neighbors, the sum in the second term aggregates information from all neighboring nodes, and ReLU is the Rectified Linear Unit function defined as ${\rm ReLU}(x)=\max[0,x]$.
    The MLP used by the GINE convolution consists of two linear layers with hidden dimension $d_h$ and GELU activation: Linear($d_h \to d_h$) $\to$ GELU $\to$ Linear($d_h \to d_h$). Each convolutional layer is followed by batch normalization and a residual connection (i.e., the input node features are added to the layer output, $\boldsymbol{h}^{(n+1)}_a= \boldsymbol{h}^{(n+1)}_a+\boldsymbol{h}^{(n)}_a$).

    After the first GINE layer, edge features are updated using a two-layer MLP that takes as input the concatenated embeddings of the source node, destination node, and current edge feature: $\boldsymbol{e}_{ab}^{(1)} = \text{MLP}(\boldsymbol{h}_a^{(1)}, \boldsymbol{h}_b^{(1)}, \boldsymbol{e}_{ab}^{(0)})$, where this MLP maps from dimension $3d_h$ to $d_h$ via Linear($3d_h \to d_h$) $\to$ GELU $\to$ Linear($d_h \to d_h$). 
    \item \underline{Readout}: After the convolutional layers, the final node embeddings from all $N_p$ pulsars are concatenated into a single vector of dimension $N_p \times d_h = 7{,}680$ and passed through a three-layer MLP readout: Linear($7{,}680 \to 1{,}024$) $\to$ GELU $\to$ Dropout(0.2) $\to$ Linear($1{,}024 \to 256$) $\to$ GELU $\to$ Dropout(0.2) $\to$ Linear($256 \to 1$), producing a single logit (unnormalized log-probability) for binary classification.
\end{enumerate}

\begin{figure}[t]
\centering
\includegraphics{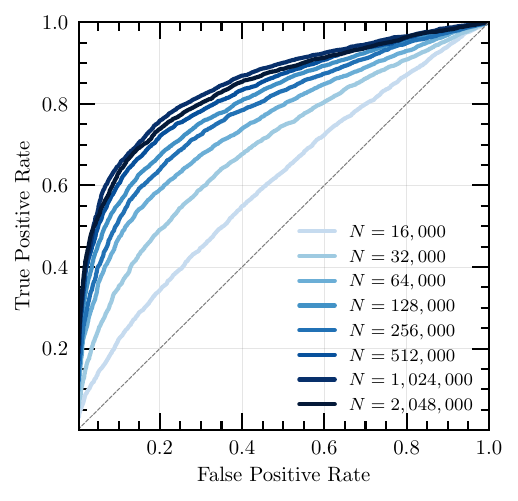}
\caption{ROC curves for the GNN classifier trained on datasets of varying size, from $N=1.6\times10^4$ to $N\simeq2\times10^6$ samples, and tested on anisotropic signals constituted by a single GWB hotspot contributing 80\% of the total GWB power in the second frequency bin.}
\label{fig:roc_nasamples}
\end{figure}
\begin{figure*}[th]
    \centering
    \includegraphics{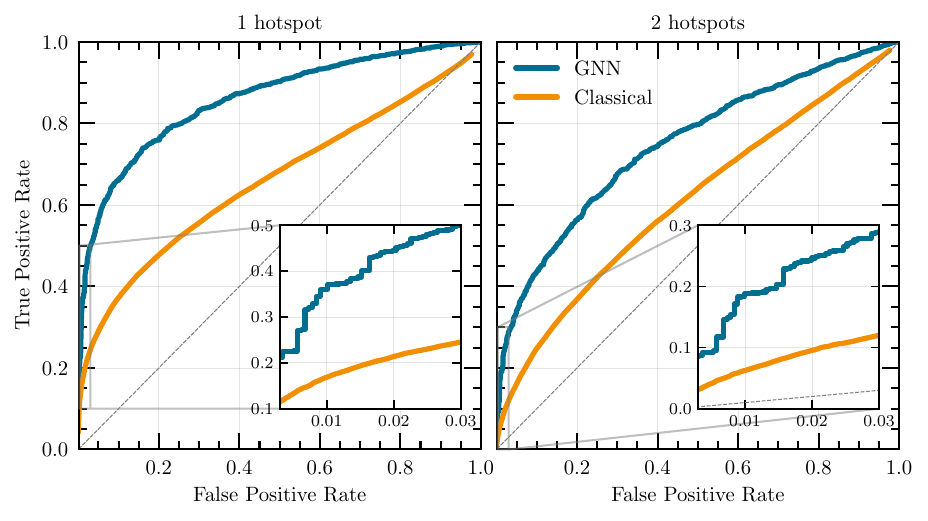}
    \caption{ROC curves for the SBI classifier (blue lines) and the classical frequentist method (orange lines). In the left panel, we report the performance of the classifier for anisotropic signals constituted by a single GWB hotspot contributing 80\% of the total GWB power in the second frequency bin versus an isotropic GWB. In the right panel, we show the results for anisotropic skies containing two GWB hotspots, each contributing 40\% of the GWB power in the second frequency bin versus an isotropic GWB. We also indicate with a blue (orange) arrow the expected rate of detections with $3\sigma$ significance for the SBI classifier (classical frequentist search). The insets show a zoomed-in version of the low false positive rate region, with the x-axis lower limit set to $3\times10^{-3}$, such that the y-axis intercept gives the expected $3\sigma$ detection rate.}
    \label{fig:roc_comp}
\end{figure*}
\subsection{Training}
\noindent Following the procedure discussed in Sec.~\ref{subsec:data}, we generate four separate datasets. The \texttt{training} dataset, consisting of approximately 4.6 million samples, is used to train the classifier.\footnote{As a reference, generating $10^5$ samples for the training dataset takes approximately 35 CPU hours, making large-scale dataset simulation feasible on standard multi-core compute nodes. Although not explored in this work, the data-generation pipeline could be re-implemented in \texttt{jax} for more efficient vectorized and parallelized simulation.} The training progress is monitored using a \texttt{validation} dataset containing approximately 0.6 million samples, and the final network performance is derived using a \texttt{test} data set of equal size. All datasets are equally split between isotropic and anisotropic GWB realizations. The number and intensity of GWB hotspots injected in the anisotropic realizations, as well as the amplitude of the isotropic GWB component, are varied across a grid of values summarized in Table~\ref{tab:datasets}.
We also generate a smaller validation subset, \texttt{test\_real}, containing 2000 samples that we use to validate some of the assumptions made in generating the training data (see the discussion in Secs.~\ref{subsubsec:residuals} and~\ref{subsubsec:noise_run}).

We train the network using Binary Cross-Entropy (BCE) loss (see App.~\ref{app:evidence_networks} for more details) with the AdamW optimizer (learning rate $10^{-3}$, weight decay $10^{-3}$) for up to 10 epochs. A ReduceLROnPlateau scheduler halves the learning rate when validation loss plateaus for 3 consecutive epochs, with early stopping after 5 epochs without improvement. We monitor training progress on a separate held-out \texttt{validation} set. Model weights are restored to the checkpoint that achieved the best validation performance.
To assess whether our results are data-limited, we train the same model on different dataset sizes, ranging from $1.6 \times 10^4$ samples to the full $5.76 \times 10^6$ samples, and evaluate the resulting classification performance. This allows us to check how performance scales with training set size. As shown in Fig.~\ref{fig:roc_nasamples}, we find that model performance saturates when the training set exceeds $10^5$ samples, suggesting we are not data-limited and that adding more training data would not yield significant improvements.

\section{Results}\label{sec:results}
\noindent In this section, we assess the capabilities of the SBI classifier developed for this work. We start by comparing its performance with that of the best-performing frequentist search strategy identified in Ref.~\cite{Konstandin:2025ifn} (i.e., a max-SNR detection statistic combined with a radiometer map parametrization, as described in Sec.~\ref{sec:review}), using the \texttt{test} dataset.
Both methods are benchmarked against two anisotropic signals: a single bright hotspot contributing 80\% of the GWB power in the second frequency bin and two bright hotspots each contributing 40\% of the power in the second bin.\footnote{For the two-hotspots case, the results of the classical methods—derived in Ref.~\cite{Konstandin:2025ifn}—placed the two hotspots at random sky locations while always keeping their angular separation fixed to $90^\circ$; in this work, we do not impose this constraint. We do not expect this difference to affect the results in any meaningful way.} 
We evaluate classifier performance using receiver operating characteristic (ROC) curves, which quantify the trade-off between detection rate (true positive rate) and false alarm rate (false positive rate) as we vary the classification threshold.
\begin{figure*}[t]
    \centering
    \includegraphics{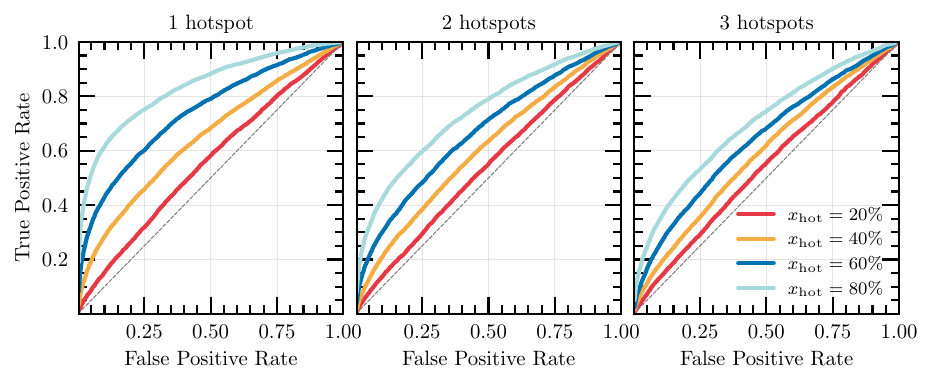}
    \caption{ROC curves for the SBI classifier for different anisotropic signals: one GWB hotspot (left panel), two GWB hotspots (central panel), three GWB hotspots (right panel). In each panel, the different lines correspond to different choices for the total contribution of the hotspots to the total GWB power: 20\% (red curve), 40\% (yellow curve), 60\% (dark blue curve), and 80\% (light blue curve).}
    \label{fig:roc_panel}
\end{figure*}

Figure~\ref{fig:roc_comp} shows the ROC curves for both methods on these two benchmark scenarios. We find that the SBI classifier significantly outperforms the classical method. For the 1-hotspot case, the expected $3\sigma$ detection rate is \one compared with 11\% for the classical frequentist method—an improvement of approximately \onegain. For the 2-hotspot case, the $3\sigma$ detection rates are \two (SBI) versus $3\%$ (classical), representing an improvement of \twogain. In deriving these results, we only used test data where the GWB amplitude was fixed to the value measured in the NANOGrav 15-year data~\cite{NANOGrav:2023gor}, i.e. $\log_{10}A_{\rm GWB}=-14.76$. 

Figure~\ref{fig:roc_panel} shows the performance of the SBI classifier as we vary the number of GWB hotspots between one and three and their total contribution to the GWB power in the second frequency bin between 20\% and 80\%. 
We assume equal-strength hotspots such that the anisotropic power is distributed uniformly among them; e.g., for two hotspots contributing 80\% of the total GWB, each contributes 40\%.
As expected, increasing the number of hotspots while keeping their total contribution to the GWB fixed decreases classifier performance, since the sky becomes increasingly isotropic.
Similarly, for a fixed number of hotspots, the classifier performance decreases as their total contribution to the GWB decreases.

\begin{figure}[t]
\centering
\includegraphics{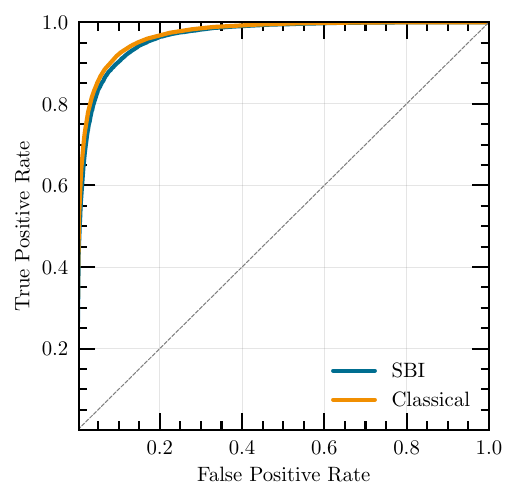}
\caption{ROC curves comparing the SBI classifier (blue) and classical frequentist method (orange) on mock cross-correlation data artificially forced to follow a Gaussian distribution.}
\label{fig:gaussian_test}
\end{figure}

Our working assumption is that the SBI classifier outperforms frequentist methods by capturing the non-Gaussian distribution of the cross-correlation estimators. To test this assumption, we train and validate both methods on a set of unrealistic data for which the cross-correlation estimators are forced to follow a multivariate normal distribution with a mean given by Eq.~\eqref{eq:orf} and a fixed covariance matrix $\boldsymbol{\Sigma}_k$\footnote{The covariance matrix for the cross-correlation coefficients depends on the GWB sky map. When generating the Gaussian dataset, we fix $\boldsymbol{\Sigma}_k$ to values corresponding to an isotropic sky, matching the assumption used in the classical frequentist likelihood.}. 
Moreover, to provide a well-defined performance target, we fix both the position and intensity of the GWB hotspot when generating these data. This allows the classical method to perform a likelihood ratio test with fully specified hypotheses: null (isotropic sky) versus alternative (known hotspot configuration). By the Neyman-Pearson Lemma, the likelihood ratio test is the optimal detection statistic in this setting and should strictly outperform the SBI classifier.\footnote{When the hotspot location and amplitude are unknown and must be reconstructed from the data, the classical frequentist method is no longer guaranteed to be optimal by the Neyman-Pearson Lemma. In this case, we find that while performance remains comparable, the SBI classifier can outperform the classical method even when cross-correlations follow a Gaussian distribution.}
The results of this test are shown in Fig.~\ref{fig:gaussian_test}. From this figure, we see that when cross-correlation estimators follow a Gaussian distribution, both approaches achieve comparable performance. This result validates our assumption that the SBI classifier's advantage stems from its ability to capture the non-Gaussian structure of the estimators, which is lost in the classical approach. Notice that for this specific test, we use a simpler network architecture consisting of an MLP with two hidden layers for the classifier, since the GNN tends to overfit on this smaller dataset with its simpler cross-correlation distribution.

A possible concern is that the specific examples of GWB anisotropies used in the training data could induce an implicit inductive bias in the classifier and reduce its sensitivity to GWB anisotropies that it never encountered during the training process. To check the amount of inductive bias introduced by the specific choice of GWB anisotropies in our training set, we tested the performance of a classifier trained only on GWB anisotropies constituted by a single hotspot containing 80\% of the power. We then evaluated this specialized classifier on test data containing signals with different hotspot configurations: multiple hotspots ($n_{\rm hot} = 2, 3$) and varying power fractions ($x_{\rm hot} = 0.4, 0.8$). Despite never encountering these configurations during training, the classifier maintained robust performance, with ROC curves comparable to those of the fully-trained classifier shown in Fig.~\ref{fig:roc_panel}.

Finally, it should be noted that, in estimating the performance of both the SBI classifier and classical frequentist search, we are analyzing only the frequency bin at which the signal was injected. Therefore, all the detection significance measures reported in this section should be understood as local significances. In a realistic analysis, we would have to search for signals in multiple frequency bins, which would introduce a trials factor that reduces the global significance~\cite{Konstandin:2025ifn}.

\subsection{Assumptions and validation}
\noindent To make the generation of $\mathcal{O}(10^6)$ training samples feasible, we had to resort to some simplifying assumptions. In this subsection, we discuss these assumptions and, where possible, explicitly check their impact on our results. The three main assumptions made in generating the training data are (see Sec.~\ref{subsec:data} for a detailed discussion of the data generation procedure and associated approximations):
\begin{itemize}
    \item In deriving the cross-correlation estimators from the mock timing data, we assumed perfect knowledge of the noise parameters. In a real measurement, these noise parameters would be unknown and would need to be inferred from the data. Typically, a preliminary Bayesian analysis modeling the GWB as a CURN process is performed to obtain posterior distributions for these quantities. These posteriors are then used to marginalize over noise parameter uncertainties by computing cross-correlation estimators and associated detection statistics over multiple random draws from the noise posteriors.
    \item We do not perform a full timing model fit of the timing residuals after injecting noise and signal in the data. Instead, we marginalize over linear deviations from the true timing model parameters in the derivation of the cross-correlation estimators~\cite{NANOGrav:2023icp} (for more details, see Sec.~\ref{subsubsec:cross_corr}).
    \item The statistical properties of the noise and signals injected into the data perfectly match those assumed in our likelihood model. In real PTA data, there will inevitably be deviations between the assumed noise models and the actual noise processes, as well as unmodeled effects such as timing glitches and instrumental systematics. 
\end{itemize}
All these approximations also apply to the mock data needed to calibrate classical frequentist methods and are unlikely to introduce a systematic bias favoring the SBI classifier in our comparison. Nevertheless, we explicitly verify that the classifier's performance remains robust when tested on more realistic data that relaxes some of these simplifying assumptions.
\begin{figure}[t]
\centering
\includegraphics{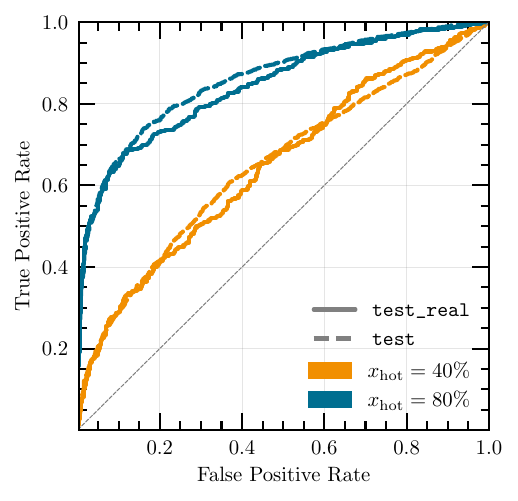}
\caption{ROC curves for the SBI classifier tested on mock data generated with (solid lines) and without (dashed lines) the inclusions of noise-marginalization and a full timing-model fit. Orange and blue lines correspond to different anisotropic signal assumptions: a single hotspot contributing 80\% of the GWB power in the second frequency bin (blue) and a single hotspot contributing 40\% (orange).}
\label{fig:roc_nmos}
\end{figure}

To do this, we use the \texttt{test\_real} dataset that drops the first two assumptions mentioned above and more closely follows the procedure used in analyzing real data (see the discussion at the end of Sec.~\ref{subsec:data} for more details). The results of this test are summarized in Fig.~\ref{fig:roc_nmos}. From this figure, we see that the classifier's performance is not significantly degraded when applied to this more realistic dataset. It is worth noting that, when using noise-marginalized data, we have multiple classifier probabilities for each realization of the timing residuals, since each choice of the noise parameters gives a different set of cross-correlation estimators (see discussion in Sec.~\ref{subsec:data}). We aggregate these by taking the median classifier probability across noise draws for each data realization, which provides a noise-averaged detection statistic while naturally accounting for noise parameter uncertainties.

While we can explicitly assess the impact of the first two assumptions using \texttt{test\_real}, the third assumption remains challenging to test using simulated data. A possibility would be to validate the classifier using sky scrambles of real PTA data, where pulsar positions are randomly rotated to destroy anisotropic correlations while maintaining realistic noise and systematics. We defer this analysis to an upcoming work.

\section{Conclusions}\label{sec:conclusions}
\noindent In this work, we have demonstrated that the Gaussian assumption underlying classical frequentist anisotropy searches significantly limits their sensitivity and introduced a simulation-based inference framework to address this limitation. Our approach trains a neural network classifier on synthetic data to learn a detection statistic directly from cross-correlation estimators, without requiring approximate Gaussian likelihoods or intermediate sky map reconstruction. 

Using a benchmark strategy similar to that developed in Ref.~\cite{Konstandin:2025ifn}, we find that this SBI classifier significantly outperforms the frequentist methods adopted in anisotropy searches by the NANOGrav~\cite{NANOGrav:2023tcn} and MeerKAT~\cite{Grunthal:2024sor} collaborations, delivering approximately \onegain (\twogain) improvements in $3\sigma$ detection rates for single (double) hotspot anisotropies.
These improvements primarily stem from the classifier's ability to capture the non-Gaussian structure of cross-correlation estimators.
We validate this interpretation by showing that both methods perform equally well when estimators are artificially constrained to follow a Gaussian distribution. Furthermore, while the classifier was trained on data generated under simplifying assumptions (perfect knowledge of noise parameters and no timing model refitting), we have verified that its performance remains robust when applied to more realistic validation datasets that include noise parameter marginalization and full timing model fits. 

While our results represent a promising step toward an optimal and fast detection strategy for GWB anisotropies, several directions remain for future work.
The classifier could be improved in several ways. First, the classifier developed in this work analyzes only a single frequency bin at a time; extending it to perform joint multi-frequency classification would enable it to exploit frequency correlations present in the data. Second, the network architecture presented here serves as a proof of principle; further optimization of the architecture and training procedure may yield additional performance gains. Third, while our current implementation focuses solely on hypothesis testing, it could be extended to perform map reconstruction. Finally, training classifiers directly on timing residuals rather than compressed cross-correlation estimators could preserve phase and polarization information, though at increased computational cost.

Beyond these improvements, further testing and validation are needed. First, we plan to test the classifier on realistic simulations of SMBHB populations, both to assess its performance on more complex anisotropy patterns and to update previous detection forecasts for SMBHB-generated GWBs (see, for example, Ref.~\cite{Lemke:2024cdu}). Second, applying the classifier to real PTA datasets would provide important insight into how unmodeled noise sources, timing glitches, and instrumental systematics affect performance. Finally, several new frequentist search strategies have been recently proposed (see, for example, Refs.~\cite{Moreschi:2025qtm, Cusin:2025xle, Agarwal:2026nxa, Curylo:2026fft}), and while in this work we focused on methods previously adopted in collaboration searches, benchmarking our SBI classifier against these newer approaches would also be valuable.

\bigskip 

\acknowledgments
\noindent The authors thank Bjorn Larsen and Joe Romano for helpful comments on the draft. AM acknowledges support from a Royal Society University Research Fellowship (URF-R1-251896). AM acknowledges the hospitality of DESY, Hamburg, where a large part of this work was completed. AL and AM are members of the NANOGrav Collaboration. NANOGrav is supported by NSF Physics Frontier Center award \#2020265. This work used the Maxwell computational resources operated at Deutsches Elektronen-Synchrotron DESY, Hamburg (Germany). JA is supported by a fellowship from the Kavli Foundation. The work of MP is supported by the Comunidad de Madrid under the Programa de Atracción de Talento Investigador with number 2024-T1TEC-3134. MP acknowledges the hospitality of Imperial College London, which provided office space during parts of this project.
TK and AL acknowledge support by the Deutsche Forschungsgemeinschaft (DFG, German Research Foundation) under Germany’s Excellence Strategy – EXC 2121 ``Quantum Universe'' – 390833306. AL thanks the Gravitational Physics Group at ETH Zurich for support, hospitality and helpful discussions during the final parts of this project.
\bigskip

\appendix
\section{Classification as Bayesian Model Comparison}
\label{app:evidence_networks}

\noindent In this appendix, we outline the mathematical justification for using a standard neural network classifier to estimate the Bayes factor, as presented in~\cite{Jeffrey:2023stk}. 

Consider a dataset $\mathbf{x}$ (in our case, the cross-correlation estimators $\hat{\rho}_{ab,k}$) and two competing models: $\mathcal{H}_0$ representing the null hypothesis (an isotropic GWB) and $\mathcal{H}_1$ representing the alternative hypothesis (an anisotropic GWB). Then, the Bayes factor $\mathcal{K}$, which acts as our detection statistic, is defined as the ratio of the model evidences:
\begin{equation}
    \mathcal{K}(\mathbf{x}) \equiv \frac{p(\mathbf{x} | \mathcal{H}_0)}{p(\mathbf{x} | \mathcal{H}_1)}.
    \label{eq:bayes_factor_def}
\end{equation}
To compute this without evaluating the intractable marginal likelihoods, we train a neural network $\rho_\phi(\mathbf{x}) = \sigma(f_\phi(\mathbf{x}))$ to perform binary classification, where $\sigma(u) = (1 + \exp(-u))^{-1}$ is the sigmoid function. This network $\rho_\phi$ takes the data $\mathbf{x}$ as input and outputs a value between 0 and 1. We provide the network with simulated data drawn from the two data distributions $p(\mathbf{x} | \mathcal{H}_0)$ and $p(\mathbf{x} | \mathcal{H}_1)$.

The network is trained to minimize the standard Binary Cross-Entropy (BCE) loss. The global optimization objective (the expected loss over all possible data realizations) is given by the functional:
\begin{align}
    \mathcal{L}[\rho_\phi] &= -\int{\mathrm{d}\mathbf{x} \, \, \bigg[p(\mathbf{x} | \mathcal{H}_0) \log \rho_\phi(\mathbf{x})} \nonumber \\
    &\qquad \qquad +\, p(\mathbf{x} | \mathcal{H}_1) \log [1 - \rho_\phi(\mathbf{x})]\bigg]
\end{align}
Minimizing this functional with respect to $\rho_\phi(\mathbf{x})$ ($\delta \mathcal{L}[\rho_\phi] / \delta \rho_\phi = 0$), we find that the optimal network $\rho^\star_\phi$ satisfies:
\begin{equation}
    \frac{p(\mathbf{x} | \mathcal{H}_0)}{\rho^\star_\phi(\mathbf{x})} - \frac{p(\mathbf{x} | \mathcal{H}_1)}{1 - \rho^\star_\phi(\mathbf{x})} = 0.
\end{equation}
Solving for $\rho^\star_\phi(\mathbf{x})$, we see that a perfectly trained, globally optimal network is directly related to the Bayes factor $\mathcal{K}$ via:
\begin{equation}
    \rho^*_\phi(\mathbf{x}) = \frac{\mathcal{K}(\mathbf{x})}{1 + \mathcal{K}(\mathbf{x})}.
\end{equation}
We can finally use the properties of the sigmoid function $\sigma$ to demonstrate that the optimal network $f_\phi^\star(\mathbf{x})$ is simply the log-Bayes factor:
\begin{equation}
    f_\phi^\star(\mathbf{x}) = \log \mathcal{K}(\mathbf{x}) = \log \frac{p(\mathbf{x} | \mathcal{H}_0)}{p(\mathbf{x} | \mathcal{H}_1)}.
\end{equation}
As such, by simply training a standard binary classifier to converge on forward-modeled mock data, we are able to directly access the (log-)Bayes factor $\mathcal{K}(\mathbf{x})$ via our neural network estimator.

\bibliographystyle{utphys}
\bibliography{ref}

@article{Hellings:1983fr,
  author  = {Hellings, R. W. and Downs, G. S.},
  title   = {{Upper Limits on the Isotropic Gravitational Radiation Background from Pulsar Timing Analysis}},
  doi     = {10.1086/183954},
  journal = {Astrophys. J. Lett.},
  volume  = {265},
  pages   = {L39--L42},
  year    = {1983}
}

@article{Konstandin:2025ifn,
    author = "Konstandin, Thomas and Lemke, Anna-Malin and Mitridate, Andrea and Perboni, Enrico",
    title = "{Prospects and limitations of PTAs anisotropy searches {\textemdash} the frequentist case}",
    eprint = "2509.07074",
    archivePrefix = "arXiv",
    primaryClass = "astro-ph.CO",
    doi = "10.1088/1475-7516/2026/02/084",
    journal = "JCAP",
    volume = "02",
    pages = "084",
    year = "2026"
}

@article{Gersbach:2024hcc,
  author        = {Gersbach, Kyle A. and Taylor, Stephen R. and Meyers, Patrick M. and Romano, Joseph D.},
  title         = {{Spatial and spectral characterization of the gravitational-wave background with the PTA optimal statistic}},
  eprint        = {2406.11954},
  archiveprefix = {arXiv},
  primaryclass  = {astro-ph.IM},
  doi           = {10.1103/PhysRevD.111.023027},
  journal       = {Phys. Rev. D},
  volume        = {111},
  number        = {2},
  pages         = {023027},
  year          = {2025}
}

@article{Gersbach:2025mhj,
    author = "Gersbach, Kyle A. and Taylor, Stephen R. and B{\'e}csy, Bence and Lemke, Anna-Malin and Mitridate, Andrea and Pol, Nihan",
    title = "{Mapping the gravitational-wave background across the spectrum with a next-generation anisotropic per-frequency optimal statistic}",
    eprint = "2509.07090",
    archivePrefix = "arXiv",
    primaryClass = "astro-ph.IM",
    doi = "10.1103/d55k-75p9",
    journal = "Phys. Rev. D",
    volume = "113",
    number = "10",
    pages = "103031",
    year = "2026"
}

@article{NANOGrav:2023gor,
  title         = {The {{NANOGrav}} 15-Year {{Data Set}}: {{Evidence}} for a {{Gravitational-Wave Background}}},
  shorttitle    = {The {{NANOGrav}} 15-Year {{Data Set}}},
  author        = {Agazie, Gabriella and others},
  year          = {2023},
  journal       = {Astrophys. J. Lett.},
  month         = jun,
  collaboration = {NANOGrav},
  volume        = {951},
  number        = {1},
  eprint        = {2306.16213},
  primaryclass  = {astro-ph.HE},
  pages         = {L8},
  doi           = {10.3847/2041-8213/acdac6},
  urldate       = {2023-08-18},
  archiveprefix = {arXiv}
}

@article{NANOGrav:2023tcn,
    author = "Agazie, Gabriella and others",
    collaboration = "NANOGrav",
    title = "{The NANOGrav 15 yr Data Set: Search for Anisotropy in the Gravitational-wave Background}",
    eprint = "2306.16221",
    archivePrefix = "arXiv",
    primaryClass = "astro-ph.HE",
    doi = "10.3847/2041-8213/acf4fd",
    journal = "Astrophys. J. Lett.",
    volume = "956",
    number = "1",
    pages = "L3",
    year = "2023"
}

@article{Reardon:2023gzh,
  title         = {Search for an Isotropic Gravitational-Wave Background with the {{Parkes Pulsar Timing Array}}},
  author        = {Reardon, Daniel J. and others},
  year          = {2023},
  month         = jun,
  journal       = {The Astrophysical Journal Letters},
  volume        = {951},
  number        = {1},
  eprint        = {2306.16215},
  primaryclass  = {astro-ph.HE},
  pages         = {L6},
  issn          = {2041-8205, 2041-8213},
  doi           = {10.3847/2041-8213/acdd02},
  urldate       = {2023-08-27},
  collaboration = {PPTA},
  archiveprefix = {arxiv}
}

@article{EPTA:2023fyk,
    author = "Antoniadis, J. and others",
    collaboration = "EPTA, InPTA:",
    title = "{The second data release from the European Pulsar Timing Array - III. Search for gravitational wave signals}",
    eprint = "2306.16214",
    archivePrefix = "arXiv",
    primaryClass = "astro-ph.HE",
    doi = "10.1051/0004-6361/202346844",
    journal = "Astron. Astrophys.",
    volume = "678",
    pages = "A50",
    year = "2023"
}

@article{Xu:2023wog,
  title         = {Searching for the Nano-{{Hertz}} Stochastic Gravitational Wave Background with the {{Chinese Pulsar Timing Array Data Release I}}},
  author        = {Xu, Heng and others},
  year          = {2023},
  month         = jun,
  journal       = {Research in Astronomy and Astrophysics},
  volume        = {23},
  number        = {7},
  collaboration = {CPTA},
  eprint        = {2306.16216},
  primaryclass  = {astro-ph.HE},
  pages         = {075024},
  issn          = {1674-4527},
  doi           = {10.1088/1674-4527/acdfa5},
  urldate       = {2023-08-27},
  archiveprefix = {arXiv}
}

@article{EPTA:2023sfo,
  title         = {The Second Data Release from the {{European Pulsar Timing Array I}}. {{The}} Dataset and Timing Analysis},
  author        = {Antoniadis, J. and others},
  year          = {2023},
  month         = jun,
  journal       = {Astronomy \& Astrophysics},
  eprint        = {2306.16224},
  primaryclass  = {astro-ph.HE},
  issn          = {0004-6361, 1432-0746},
  doi           = {10.1051/0004-6361/202346841},
  urldate       = {2023-08-27},
  archiveprefix = {arxiv},
  collaboration = "EPTA, InPTA"
}

@article{NANOGrav:2023hvm,
    author = "Afzal, Adeela and others",
    collaboration = "NANOGrav",
    title = "{The NANOGrav 15 yr Data Set: Search for Signals from New Physics}",
    eprint = "2306.16219",
    archivePrefix = "arXiv",
    primaryClass = "astro-ph.HE",
    reportNumber = "FERMILAB-PUB-23-589-T",
    doi = "10.3847/2041-8213/acdc91",
    journal = "Astrophys. J. Lett.",
    volume = "951",
    number = "1",
    pages = "L11",
    year = "2023"
}

@article{NANOGrav:2023hfp,
  author        = {Agazie, Gabriella and others},
  collaboration = {NANOGrav},
  title         = {{The NANOGrav 15 yr Data Set: Constraints on Supermassive Black Hole Binaries from the Gravitational-wave Background}},
  eprint        = {2306.16220},
  archiveprefix = {arXiv},
  primaryclass  = {astro-ph.HE},
  doi           = {10.3847/2041-8213/ace18b},
  journal       = {Astrophys. J. Lett.},
  volume        = {952},
  number        = {2},
  pages         = {L37},
  year          = {2023}
}

@article{Taylor:2013esa,
  title         = {Searching {{For Anisotropic Gravitational-wave Backgrounds Using Pulsar Timing Arrays}}},
  author        = {Taylor, Stephen R. and Gair, Jonathan R.},
  year          = {2013},
  month         = oct,
  journal       = {Physical Review D},
  volume        = {88},
  number        = {8},
  eprint        = {1306.5395},
  primaryclass  = {gr-qc},
  pages         = {084001},
  issn          = {1550-7998, 1550-2368},
  doi           = {10.1103/PhysRevD.88.084001},
  urldate       = {2023-09-30},
  archiveprefix = {arxiv}
}

@article{Gardiner:2023zzr,
    author = "Gardiner, Emiko C. and Kelley, Luke Zoltan and Lemke, Anna-Malin and Mitridate, Andrea",
    title = "{Beyond the Background: Gravitational-wave Anisotropy and Continuous Waves from Supermassive Black Hole Binaries}",
    eprint = "2309.07227",
    archivePrefix = "arXiv",
    primaryClass = "astro-ph.HE",
    doi = "10.3847/1538-4357/ad2be8",
    journal = "Astrophys. J.",
    volume = "965",
    number = "2",
    pages = "164",
    year = "2024"
}

@article{Lemke:2024cdu,
    author = "Lemke, Anna-Malin and Mitridate, Andrea and Gersbach, Kyle A.",
    title = "{Detecting gravitational wave anisotropies from supermassive black hole binaries}",
    eprint = "2407.08705",
    archivePrefix = "arXiv",
    primaryClass = "astro-ph.HE",
    reportNumber = "DESY-24-097",
    doi = "10.1103/PhysRevD.111.063068",
    journal = "Phys. Rev. D",
    volume = "111",
    number = "6",
    pages = "063068",
    year = "2025"
}

@article{Taylor:2015udp,
  title         = {Limits on Anisotropy in the Nanohertz Stochastic Gravitational-Wave Background},
  author        = {Taylor, S. R. and others},
  year          = {2015},
  month         = jul,
  journal       = {Physical Review Letters},
  volume        = {115},
  number        = {4},
  eprint        = {1506.08817},
  primaryclass  = {astro-ph.HE},
  pages         = {041101},
  issn          = {0031-9007, 1079-7114},
  doi           = {10.1103/PhysRevLett.115.041101},
  urldate       = {2023-12-11},
  archiveprefix = {arxiv}
}

@article{Pol:2022sjn,
  title         = {Forecasting Pulsar Timing Array Sensitivity to Anisotropy in the Stochastic Gravitational Wave Background},
  author        = {Pol, Nihan and Taylor, Stephen R. and Romano, Joseph D.},
  year          = {2022},
  month         = dec,
  journal       = {The Astrophysical Journal},
  volume        = {940},
  number        = {2},
  eprint        = {2206.09936},
  primaryclass  = {astro-ph.HE},
  pages         = {173},
  issn          = {0004-637X, 1538-4357},
  doi           = {10.3847/1538-4357/ac9836},
  urldate       = {2023-09-08},
  archiveprefix = {arxiv}
}

@article{Hazboun:2023tiq,
  author        = {Hazboun, Jeffrey S. and Meyers, Patrick M. and Romano, Joseph D. and Siemens, Xavier and Archibald, Anne M.},
  title         = {{Analytic distribution of the optimal cross-correlation statistic for stochastic gravitational-wave-background searches using pulsar timing arrays}},
  eprint        = {2305.01116},
  archiveprefix = {arXiv},
  primaryclass  = {gr-qc},
  doi           = {10.1103/PhysRevD.108.104050},
  journal       = {Phys. Rev. D},
  volume        = {108},
  number        = {10},
  pages         = {104050},
  year          = {2023}
}

@software{Vallisneri_nanograv_discovery_2025,
  author  = {Vallisneri, Michele and Meyers, Patrick M. and Wright, David and Johnson, Aaron D. and Baier, Jeremy G. and van Haasteren, Rutger},
  doi     = {10.5281/zenodo.17711453},
  license = {["MIT"]},
  month   = nov,
  title   = {{nanograv/discovery}},
  url     = {https://github.com/nanograv/discovery},
  year    = {2025}
}

@article{discovery,
  author= {Vallisneri, Michele and others},
  title = {\texttt{Discovery}: the next-generation
  pulsar-timing-array data-analysis package},
  journal= {in preparation}
 }

@article{Jeffrey:2023stk,
  author        = {Jeffrey, Niall and Wandelt, Benjamin D.},
  title         = {{Evidence Networks: simple losses for fast, amortized, neural Bayesian model comparison}},
  eprint        = {2305.11241},
  archiveprefix = {arXiv},
  primaryclass  = {cs.LG},
  doi           = {10.1088/2632-2153/ad1a4d},
  journal       = {Mach. Learn. Sci. Tech.},
  volume        = {5},
  number        = {1},
  pages         = {015008},
  year          = {2024}
}

@article{AnauMontel:2024flo,
  author        = {Anau Montel, Noemi and Alvey, James and Weniger, Christoph},
  title         = {{Tests for model misspecification in simulation-based inference: From local distortions to global model checks}},
  eprint        = {2412.15100},
  archiveprefix = {arXiv},
  primaryclass  = {astro-ph.IM},
  doi           = {10.1103/PhysRevD.111.083013},
  journal       = {Phys. Rev. D},
  volume        = {111},
  number        = {8},
  pages         = {083013},
  year          = {2025}
}

@article{NANOGrav:2023icp,
  title    = {The {NANOGrav} 15-year {Gravitational}-{Wave} {Background} {Methods}},
  volume   = {109},
  issn     = {2470-0010, 2470-0029},
  url      = {http://arxiv.org/abs/2306.16223},
  doi      = {10.1103/PhysRevD.109.103012},
  number   = {10},
  urldate  = {2026-04-23},
  journal  = {Physical Review D},
  author   = {Johnson, Aaron D. and Meyers, Patrick M. and Baker, Paul T. and Cornish, Neil J. and Hazboun, Jeffrey S. and Littenberg, Tyson B. and Romano, Joseph D. and Taylor, Stephen R. and Vallisneri, Michele and Vigeland, Sarah J. and Olum, Ken D. and Siemens, Xavier and Ellis, Justin A. and Haasteren, Rutger van and Hourihane, Sophie and Agazie, Gabriella and Anumarlapudi, Akash and Archibald, Anne M. and Arzoumanian, Zaven and Blecha, Laura and Brazier, Adam and Brook, Paul R. and Burke-Spolaor, Sarah and Bécsy, Bence and Casey-Clyde, J. Andrew and Charisi, Maria and Chatterjee, Shami and Chatziioannou, Katerina and Cohen, Tyler and Cordes, James M. and Crawford, Fronefield and Cromartie, H. Thankful and Crowter, Kathryn and DeCesar, Megan E. and Demorest, Paul B. and Dolch, Timothy and Drachler, Brendan and Ferrara, Elizabeth C. and Fiore, William and Fonseca, Emmanuel and Freedman, Gabriel E. and Garver-Daniels, Nate and Gentile, Peter A. and Glaser, Joseph and Good, Deborah C. and Gültekin, Kayhan and Jennings, Ross J. and Jones, Megan L. and Kaiser, Andrew R. and Kaplan, David L. and Kelley, Luke Zoltan and Kerr, Matthew and Key, Joey S. and Laal, Nima and Lam, Michael T. and Lamb, William G. and Lazio, T. Joseph W. and Lewandowska, Natalia and Liu, Tingting and Lorimer, Duncan R. and Lynch, Ryan S. and Ma, Chung-Pei and Madison, Dustin R. and McEwen, Alexander and McKee, James W. and McLaughlin, Maura A. and McMann, Natasha and Meyers, Bradley W. and Mingarelli, Chiara M. F. and Mitridate, Andrea and Ng, Cherry and Nice, David J. and Ocker, Stella Koch and Pennucci, Timothy T. and Perera, Benetge B. P. and Pol, Nihan S. and Radovan, Henri A. and Ransom, Scott M. and Ray, Paul S. and Sardesai, Shashwat C. and Schmiedekamp, Carl and Schmiedekamp, Ann and Schmitz, Kai and Shapiro-Albert, Brent J. and Simon, Joseph and Siwek, Magdalena S. and Stairs, Ingrid H. and Stinebring, Daniel R. and Stovall, Kevin and Susobhanan, Abhimanyu and Swiggum, Joseph K. and Turner, Jacob E. and Unal, Caner and Wahl, Haley M. and Witt, Caitlin A. and Young, Olivia},
  month    = may,
  year     = {2024},
  pages    = {103012}
}

@article{Grunthal:2024sor,
  title      = {The {MeerKAT} {Pulsar} {Timing} {Array}: {Maps} of the gravitational-wave sky with the 4.5 year data release},
  volume     = {536},
  issn       = {0035-8711, 1365-2966},
  shorttitle = {The {MeerKAT} {Pulsar} {Timing} {Array}},
  url        = {http://arxiv.org/abs/2412.01214},
  doi        = {10.1093/mnras/stae2573},
  collaboration = {MeerKAT},
  language   = {en},
  number     = {2},
  urldate    = {2025-02-17},
  journal    = {Monthly Notices of the Royal Astronomical Society},
  author     = {Grunthal, Kathrin and Nathan, Rowina S. and Thrane, Eric and Champion, David J. and Miles, Matthew T. and Shannon, Ryan M. and Kulkarni, Atharva D. and Abbate, Federico and Buchner, Sarah and Cameron, Andrew D. and Geyer, Marisa and Gitika, Pratyasha and Keith, Michael J. and Kramer, Michael and Lasky, Paul D. and Parthasarathy, Aditya and Reardon, Daniel J. and Singha, Jaikhomba and Krishnan, Vivek Venkatraman},
  month      = dec,
  year       = {2024},
  pages      = {1501--1517}
}

@misc{Curylo:2026fft,
  title     = {A comprehensive framework for phase-coherent mapping of the gravitational-wave sky with pulsar timing arrays},
  url       = {http://arxiv.org/abs/2604.19073},
  doi       = {10.48550/arXiv.2604.19073},
  urldate   = {2026-04-26},
  publisher = {arXiv},
  author    = {Curyło, Małgorzata and Thrane, Eric and Lasky, Paul D. and Gaynor, Dawson S.},
  month     = apr,
  year      = {2026},
  note      = {arXiv:2604.19073 [astro-ph]}
}

@misc{Cusin:2025xle,
  title     = {Measuring anisotropies in the {PTA} band with cross-correlations},
  url       = {http://arxiv.org/abs/2502.17401},
  doi       = {10.48550/arXiv.2502.17401},
  urldate   = {2025-02-25},
  publisher = {arXiv},
  author    = {Cusin, Giulia and Pitrou, Cyril and Pijnenburg, Martin and Sesana, Alberto},
  month     = feb,
  year      = {2025}
}

@misc{Agarwal:2026nxa,
  title     = {Addressing leakage and mode suppression in angular power spectrum estimation for gravitational-wave backgrounds using pulsar timing arrays},
  url       = {http://arxiv.org/abs/2602.20075},
  doi       = {10.48550/arXiv.2602.20075},
  urldate   = {2026-04-16},
  publisher = {arXiv},
  author    = {Agarwal, Deepali and Romano, Joseph D. and Ali-Haïmoud, Yacine and Smith, Tristan L.},
  month     = feb,
  year      = {2026}
}

@article{NANOGrav:2023hde,
    author = "Agazie, Gabriella and others",
    collaboration = "NANOGrav",
    title = "{The NANOGrav 15 yr Data Set: Observations and Timing of 68 Millisecond Pulsars}",
    eprint = "2306.16217",
    archivePrefix = "arXiv",
    primaryClass = "astro-ph.HE",
    doi = "10.3847/2041-8213/acda9a",
    journal = "Astrophys. J. Lett.",
    volume = "951",
    number = "1",
    pages = "L9",
    year = "2023"
}

@article{Zic:2023gta,
  title    = {The {Parkes} {Pulsar} {Timing} {Array} {Third} {Data} {Release}},
  volume   = {40},
  url      = {http://arxiv.org/abs/2306.16230},
  doi      = {10.1017/pasa.2023.36},
  urldate  = {2023-08-27},
  journal  = {Publ. Astron. Soc. Austral.},
  author   = {Zic, Andrew and Reardon, Daniel J. and Kapur, Agastya and Hobbs, George and Mandow, Rami and Curyło, Małgorzata and Shannon, Ryan M. and Askew, Jacob and Bailes, Matthew and Bhat, N. D. Ramesh and Cameron, Andrew and Chen, Zu-Cheng and Dai, Shi and Di Marco, Valentina and Feng, Yi and Kerr, Matthew and Kulkarni, Atharva and Lower, Marcus E. and Luo, Rui and Manchester, Richard N. and Miles, Matthew T. and Nathan, Rowina S. and Osłowski, Stefan and Rogers, Axl F. and Russell, Christopher J. and Spiewak, Renée and Thyagarajan, Nithyanandan and Toomey, Lawrence and Wang, Shuangqiang and Zhang, Lei and Zhang, Songbo and Zhu, Xing-Jiang},
  month    = jul,
  collaboration = {PPTA},
  year     = {2023},
  pages    = {e049}
}

@misc{justin_ellis_2017_251456,
  author = {Justin Ellis and
            Rutger van Haasteren},
  title  = {jellis18/PAL2: PAL2},
  month  = jan,
  year   = 2017,
  doi    = {10.5281/zenodo.251456},
  url    = {https://doi.org/10.5281/zenodo.251456}
}

@misc{pta_replicator,
  author       = {Bence {B{\'e}csy} and Jeff Hazboun and Aaron Johnson},
  title        = {pta\_replicator},
  year         = {2025},
  howpublished = {GitHub repository},
  note         = {Available at \url{https://github.com/bencebecsy/pta_replicator}},
  url          = {https://github.com/bencebecsy/pta_replicator}
}

@article{EPTA:2023akd,
    author = "Antoniadis, J. and others",
    collaboration = "EPTA, InPTA",
    title = "{The second data release from the European Pulsar Timing Array - II. Customised pulsar noise models for spatially correlated gravitational waves}",
    eprint = "2306.16225",
    archivePrefix = "arXiv",
    primaryClass = "astro-ph.HE",
    doi = "10.1051/0004-6361/202346842",
    journal = "Astron. Astrophys.",
    volume = "678",
    pages = "A49",
    year = "2023"
}

@article{NANOGrav:2023ctt,
  title      = {The {NANOGrav} 15-{Year} {Data} {Set}: {Detector} {Characterization} and {Noise} {Budget}},
  volume     = {951},
  issn       = {2041-8205, 2041-8213},
  shorttitle = {The {NANOGrav} 15-{Year} {Data} {Set}},
  url        = {http://arxiv.org/abs/2306.16218},
  doi        = {10.3847/2041-8213/acda88},
  number     = {1},
  urldate    = {2023-08-18},
  journal    = {The Astrophysical Journal Letters},
  author     = {Agazie, Gabriella and Anumarlapudi, Akash and Archibald, Anne M. and Arzoumanian, Zaven and Baker, Paul T. and Bécsy, Bence and Blecha, Laura and Brazier, Adam and Brook, Paul R. and Burke-Spolaor, Sarah and Charisi, Maria and Chatterjee, Shami and Cohen, Tyler and Cordes, James M. and Cornish, Neil J. and Crawford, Fronefield and Cromartie, H. Thankful and Crowter, Kathryn and Decesar, Megan E. and Demorest, Paul B. and Dolch, Timothy and Drachler, Brendan and Ferrara, Elizabeth C. and Fiore, William and Fonseca, Emmanuel and Freedman, Gabriel E. and Garver-Daniels, Nate and Gentile, Peter A. and Glaser, Joseph and Good, Deborah C. and Guertin, Lydia and Gültekin, Kayhan and Hazboun, Jeffrey S. and Jennings, Ross J. and Johnson, Aaron D. and Jones, Megan L. and Kaiser, Andrew R. and Kaplan, David L. and Kelley, Luke Zoltan and Kerr, Matthew and Key, Joey S. and Laal, Nima and Lam, Michael T. and Lamb, William G. and Lazio, T. Joseph W. and Lewandowska, Natalia and Liu, Tingting and Lorimer, Duncan R. and Luo, Jing and Lynch, Ryan S. and Ma, Chung-Pei and Madison, Dustin R. and Mcewen, Alexander and Mckee, James W. and Mclaughlin, Maura A. and Mcmann, Natasha and Meyers, Bradley W. and Mingarelli, Chiara M. F. and Mitridate, Andrea and Ng, Cherry and Nice, David J. and Ocker, Stella Koch and Olum, Ken D. and Pennucci, Timothy T. and Perera, Benetge B. P. and Pol, Nihan S. and Radovan, Henri A. and Ransom, Scott M. and Ray, Paul S. and Romano, Joseph D. and Sardesai, Shashwat C. and Schmiedekamp, Ann and Schmiedekamp, Carl and Schmitz, Kai and Shapiro-Albert, Brent J. and Siemens, Xavier and Simon, Joseph and Siwek, Magdalena S. and Stairs, Ingrid H. and Stinebring, Daniel R. and Stovall, Kevin and Susobhanan, Abhimanyu and Swiggum, Joseph K. and Taylor, Stephen R. and Turner, Jacob E. and Unal, Caner and Vallisneri, Michele and Vigeland, Sarah J. and Wahl, Haley M. and Witt, Caitlin A. and Young, Olivia},
  month      = jun,
  collaboration = {NANOGrav},
  year       = {2023},
  pages      = {L10}
}

@article{Miles:2024rjc,
    author = "Miles, Matthew T. and others",
    title = "{The MeerKAT Pulsar Timing Array: the 4.5-yr data release and the noise and stochastic signals of the millisecond pulsar population}",
    eprint = "2412.01148",
    archivePrefix = "arXiv",
    primaryClass = "astro-ph.HE",
    doi = "10.1093/mnras/stae2572",
    journal = "Mon. Not. Roy. Astron. Soc.",
    volume = "536",
    number = "2",
    pages = "1467--1488",
    collaboration = {MeerKAT},
    year = "2024"
}

@article{Moreschi:2025qtm,
    author = "Moreschi, Beatrice Eleonora and Valtolina, Serena and Sesana, Alberto and Shaifullah, Golam and Falxa, Mikel and Speri, Lorenzo and Izquierdo-Villalba, David and Chalumeau, Aurelien",
    title = "{Dissecting the nanoHz gravitational wave sky: frequency-correlated anisotropy induced by eccentric supermassive black hole binaries}",
    eprint = "2506.14882",
    archivePrefix = "arXiv",
    primaryClass = "astro-ph.GA",
    month = "6",
    year = "2025"
}

@article{Vigeland:2018ipb,
    author = "Vigeland, Sarah J. and Islo, Kristina and Taylor, Stephen R. and Ellis, Justin A.",
    title = "{Noise-marginalized optimal statistic: A robust hybrid frequentist-Bayesian statistic for the stochastic gravitational-wave background in pulsar timing arrays}",
    eprint = "1805.12188",
    archivePrefix = "arXiv",
    primaryClass = "astro-ph.IM",
    doi = "10.1103/PhysRevD.98.044003",
    journal = "Phys. Rev. D",
    volume = "98",
    pages = "044003",
    year = "2018"
}

@article{Cranmer:2019eaq,
    title = {The frontier of simulation-based inference},
    volume = {117},
    issn = {0027-8424, 1091-6490},
    url = {http://arxiv.org/abs/1911.01429},
    doi = {10.1073/pnas.1912789117},
    number = {48},
    urldate = {2026-05-01},
    journal = {Proceedings of the National Academy of Sciences},
    author = {Cranmer, Kyle and Brehmer, Johann and Louppe, Gilles},
    month = dec,
    year = {2020},
    note = {arXiv:1911.01429 [stat]},
    pages = {30055--30062},
}

@article{Neyman:1933wgr,
    title = {On the {Problem} of the {Most} {Efficient} {Tests} of {Statistical} {Hypotheses}},
    volume = {231},
    doi = {10.1098/rsta.1933.0009},
    number = {694-706},
    journal = {Phil. Trans. Roy. Soc. Lond. A},
    author = {Neyman, Jerzy and Pearson, Egon Sharpe},
    year = {1933},
    pages = {289--337},
}

@article{Caprini:2018mtu,
    title = {Cosmological {Backgrounds} of {Gravitational} {Waves}},
    volume = {35},
    issn = {0264-9381, 1361-6382},
    url = {http://arxiv.org/abs/1801.04268},
    doi = {10.1088/1361-6382/aac608},
    number = {16},
    urldate = {2023-08-31},
    journal = {Classical and Quantum Gravity},
    author = {Caprini, Chiara and Figueroa, Daniel G.},
    month = jul,
    year = {2018},
    pages = {163001},
}

@article{LISACosmologyWorkingGroup:2022kbp,
    title = {Probing {Anisotropies} of the {Stochastic} {Gravitational} {Wave} {Background} with {LISA}},
    volume = {11},
    issn = {1475-7516},
    url = {http://arxiv.org/abs/2201.08782},
    doi = {10.1088/1475-7516/2022/11/009},
    number = {11},
    urldate = {2023-09-10},
    journal = {JCAP},
    author = {Bartolo, Nicola and Bertacca, Daniele and Caldwell, Robert and Contaldi, Carlo R. and Cusin, Giulia and De Luca, Valerio and Dimastrogiovanni, Emanuela and Fasiello, Matteo and Figueroa, Daniel G. and Franciolini, Gabriele and Jenkins, Alexander C. and Peloso, Marco and Pieroni, Mauro and Renzini, Arianna and Ricciardone, Angelo and Riotto, Antonio and Sakellariadou, Mairi and Sorbo, Lorenzo and Tasinato, Gianmassimo and Torrado, Jesus and Clesse, Sebastien and Kuroyanagi, Sachiko},
    month = nov,
    year = {2022},
    pages = {009},
}

@article{Mingarelli:2013dsa,
    title = {Characterising gravitational wave stochastic background anisotropy with {Pulsar} {Timing} {Arrays}},
    volume = {88},
    issn = {1550-7998, 1550-2368},
    url = {http://arxiv.org/abs/1306.5394},
    doi = {10.1103/PhysRevD.88.062005},
    number = {6},
    urldate = {2023-09-30},
    journal = {Physical Review D},
    author = {Mingarelli, Chiara M. F. and Sidery, Trevor and Mandel, Ilya and Vecchio, Alberto},
    month = sep,
    year = {2013},
    pages = {062005},
}

@article{Becsy:2022zbu,
    author = "B{\'e}csy, Bence and Cornish, Neil J. and Digman, Matthew C.",
    title = "{Fast Bayesian analysis of individual binaries in pulsar timing array data}",
    eprint = "2204.07160",
    archivePrefix = "arXiv",
    primaryClass = "gr-qc",
    doi = "10.1103/PhysRevD.105.122003",
    journal = "Phys. Rev. D",
    volume = "105",
    number = "12",
    pages = "122003",
    year = "2022"
}

@ARTICLE{2016arXiv160608415H,
       author = {{Hendrycks}, Dan and {Gimpel}, Kevin},
        title = "{Gaussian Error Linear Units (GELUs)}",
      journal = {arXiv e-prints},
         year = 2016,
        month = jun,
          eid = {arXiv:1606.08415},
        pages = {arXiv:1606.08415},
          doi = {10.48550/arXiv.1606.08415},
archivePrefix = {arXiv},
       eprint = {1606.08415},
 primaryClass = {cs.LG},
       adsurl = {https://ui.adsabs.harvard.edu/abs/2016arXiv160608415H}
}

@ARTICLE{2019arXiv190512265H,
       author = {{Hu}, Weihua and {Liu}, Bowen and {Gomes}, Joseph and {Zitnik}, Marinka and {Liang}, Percy and {Pande}, Vijay and {Leskovec}, Jure},
        title = "{Strategies for Pre-training Graph Neural Networks}",
      journal = {arXiv e-prints},
         year = 2019,
        month = may,
          eid = {arXiv:1905.12265},
        pages = {arXiv:1905.12265},
          doi = {10.48550/arXiv.1905.12265},
archivePrefix = {arXiv},
       eprint = {1905.12265},
 primaryClass = {cs.LG},
       adsurl = {https://ui.adsabs.harvard.edu/abs/2019arXiv190512265H}
}

@article{Hobbs:2006cd,
    title = {tempo2, a new pulsar-timing package - {I}. {An} overview: tempo2, a new pulsar-timing package - {I}. {Overview}},
    volume = {369},
    issn = {00358711},
    shorttitle = {tempo2, a new pulsar-timing package - {I}. {An} overview},
    url = {http://mnras.oxfordjournals.org/cgi/doi/10.1111/j.1365-2966.2006.10302.x},
    doi = {10.1111/j.1365-2966.2006.10302.x},
    language = {en},
    number = {2},
    urldate = {2023-09-15},
    journal = {Monthly Notices of the Royal Astronomical Society},
    author = {Hobbs, G. B. and Edwards, R. T. and Manchester, R. N.},
    year = {2006},
    pages = {655--672},
}

@BOOK{2004hpa..book.....L,
       author = {{Lorimer}, D.~R. and {Kramer}, M.},
        title = "{Handbook of Pulsar Astronomy}",
         year = 2004,
       volume = {4},
       adsurl = {https://ui.adsabs.harvard.edu/abs/2004hpa..book.....L}
}

@article{Taylor:1993an,
    author = "Taylor, Joseph H.",
    title = "{Pulsar timing and relativistic gravity}",
    doi = "10.1088/0264-9381/10/S/017",
    journal = "Class. Quant. Grav.",
    volume = "10",
    pages = "S167--S174",
    year = "1993"
}

@article{EPTA:2023xxk,
    author = "Antoniadis, J. and others",
    collaboration = "EPTA, InPTA",
    title = "{The second data release from the European Pulsar Timing Array - IV. Implications for massive black holes, dark matter, and the early Universe}",
    eprint = "2306.16227",
    archivePrefix = "arXiv",
    primaryClass = "astro-ph.CO",
    doi = "10.1051/0004-6361/202347433",
    journal = "Astron. Astrophys.",
    volume = "685",
    pages = "A94",
    year = "2024"
}

@article{Depta:2024ykq,
    author = "Depta, Paul Frederik and Domcke, Valerie and Franciolini, Gabriele and Pieroni, Mauro",
    title = "{Pulsar timing array sensitivity to anisotropies in the gravitational wave background}",
    eprint = "2407.14460",
    archivePrefix = "arXiv",
    primaryClass = "astro-ph.CO",
    reportNumber = "CERN-TH-2024-116",
    doi = "10.1103/PhysRevD.111.083039",
    journal = "Phys. Rev. D",
    volume = "111",
    number = "8",
    pages = "083039",
    year = "2025"
}

@article{Domcke:2025esw,
    author = "Domcke, Valerie and Franciolini, Gabriele and Pieroni, Mauro",
    title = "{Cosmic Variance in Anisotropy Searches at Pulsar Timing Arrays}",
    eprint = "2508.21131",
    archivePrefix = "arXiv",
    primaryClass = "astro-ph.CO",
    reportNumber = "CERN-TH-2025-159",
    month = "8",
    year = "2025"
}

@article{Konstandin:2024fyo,
    author = "Konstandin, Thomas and Lemke, Anna-Malin and Mitridate, Andrea and Perboni, Enrico",
    title = "{The impact of cosmic variance on PTAs anisotropy searches}",
    eprint = "2408.07741",
    archivePrefix = "arXiv",
    primaryClass = "astro-ph.CO",
    reportNumber = "DESY-24-118",
    doi = "10.1088/1475-7516/2025/04/059",
    journal = "JCAP",
    volume = "04",
    pages = "059",
    year = "2025"
}

@article{Shih:2023jme,
    author = "Shih, David and Freytsis, Marat and Taylor, Stephen R. and Dror, Jeff A. and Smyth, Nolan",
    title = "{Fast Parameter Inference on Pulsar Timing Arrays with Normalizing Flows}",
    eprint = "2310.12209",
    archivePrefix = "arXiv",
    primaryClass = "astro-ph.IM",
    doi = "10.1103/PhysRevLett.133.011402",
    journal = "Phys. Rev. Lett.",
    volume = "133",
    number = "1",
    pages = "011402",
    year = "2024"
}

@article{Vallisneri:2024xfk,
    author = "Vallisneri, Michele and Crisostomi, Marco and Johnson, Aaron D. and Meyers, Patrick M.",
    title = "{Rapid Parameter Estimation for Pulsar-Timing-Array Datasets with Variational Inference and Normalizing Flows}",
    eprint = "2405.08857",
    archivePrefix = "arXiv",
    primaryClass = "gr-qc",
    doi = "10.1103/p3f7-rbmv",
    journal = "Phys. Rev. Lett.",
    volume = "135",
    number = "7",
    pages = "071401",
    year = "2025"
}

@article{Lai:2025xov,
    author = "Lai, Junrong and Li, Changhong",
    title = "{Accelerated Bayesian inference for pulsar timing arrays: Normalizing flows for rapid model comparison across stochastic gravitational-wave background sources}",
    eprint = "2504.04211",
    archivePrefix = "arXiv",
    primaryClass = "astro-ph.CO",
    doi = "10.1103/7j5m-m9j7",
    journal = "Phys. Rev. D",
    volume = "112",
    number = "2",
    pages = "023533",
    year = "2025"
}

@article{Laal:2024trp,
    author = "Laal, Nima and others",
    title = "{Deep Neural Emulation of the Supermassive Black Hole Binary Population}",
    eprint = "2411.10519",
    archivePrefix = "arXiv",
    primaryClass = "astro-ph.IM",
    doi = "10.3847/1538-4357/adb4ef",
    journal = "Astrophys. J.",
    volume = "982",
    number = "1",
    pages = "55",
    year = "2025"
}
\end{document}